\documentclass[prb,aps,amsmath,floatfixaps,showpacs,groupedaddress]{revtex4-2}
\usepackage{graphicx,natbib,amsmath,latexsym,amssymb}
\usepackage{subcaption}
\usepackage{color}
\def\be{\begin{equation}}
\def\ee{\end{equation}}
\def\bea{\begin{eqnarray}}
\def\eea{\end{eqnarray}}
\def\bw{\begin{widetext}}
\def\ew{\end{widetext}}
\def\nn{\nonumber}
\begin{document}
\title{Pursuing equitable access to vaccines for the next epidemic}

\author{Hsin-Ju Chou}
\affiliation{Department of Physics, National Kaohsiung Normal University, Kaohsiung 82444, Taiwan, R.O.C.}
\author{Jing-Yuan Ko}
\affiliation{Department of Physics, National Kaohsiung Normal University, Kaohsiung 82444, Taiwan, R.O.C.}
\author{Sung-Po Chao}
\affiliation{Department of Physics, National Kaohsiung Normal University, Kaohsiung 82444, Taiwan, R.O.C.}
\date{\today}
\begin{abstract}
To mitigate the pandemic stemming from COVID-19, numerous nations have initiated extensive vaccination campaigns for their citizens since late 2020. While affluent countries have predominantly received vaccine allocations, fewer doses have been dispatched to nations with lower average incomes. This unequal distribution not only widens the disparity between wealthy and impoverished regions but also prolongs the pandemic, evident in the emergence of new viral variants. Our research delves into the correlation between the duration of the pandemic and the timing of vaccine distribution between two countries with migratory ties. By using a pair of coupled Susceptible-Infected-Recovered-Deceased (SIRD) models incorporating vaccination data, we demonstrate that timely sharing of vaccines benefits both nations, regardless of the presence of viral variants. This underscores that in the realm of vaccine distribution, self-interest and altruism are not mutually exclusive.
\end{abstract}
\pacs{}
\maketitle

\section{Introduction}
The COVID-19 pandemic, originating in Wuhan, China in December 2019, is caused by the highly contagious Severe Acute Respiratory Syndrome coronavirus 2 (SARS-CoV-2)\cite{wikic}. The primary modes of transmission include airborne transmission through aerosols\cite{Schooley,Wang,Lewis} or coming into contact with surfaces contaminated by the virus\cite{Cao}. This rapid spread led to a global pandemic. Initially, non-pharmaceutical interventions (NPIs) such as social distancing, surface cleaning, mask-wearing, isolation and quarantine measures, as well as restrictions on mass gatherings and travel, were implemented to curb transmission\cite{fmask,mask2,who,MUG}. While these measures helped slow the spread of COVID-19, they also had adverse effects on the economy and took a toll on mental health\cite{Ayouni,Martin,Chao}.\\

The development and widespread use of vaccines and treatments eventually brought an end to the pandemic, officially declared by the World Health Organization (WHO) on May 5, 2023\cite{wikic}. At present, daily life in most countries has returned to pre-pandemic norms. The virus is now managed similarly to seasonal influenza, with new vaccine booster shots tailored to target emerging virus variants.\\

Throughout the pandemic, the COVID-19 Vaccines Global Access (COVAX), operating as one of the four pillars within the Access to COVID-19 Tools-Accelerator (ACT-A), has been orchestrating international efforts to facilitate fair access to COVID-19 tests, treatments, and vaccines for low to middle-income countries\cite{Turner,Khairi,COVAX}. Despite these endeavors, the fair distribution of vaccine supplies has been hindered by high-income nations securing large quantities of vaccines through advance orders, leaving lower-income countries scrambling for remaining doses\cite{Khairi}. The significance of ensuring equitable access to COVID-19 vaccines has been extensively discussed across various models including multistrain compartment models\cite{Ye}, stochastic compartment models\cite{Moore,Gozzi,Li}, statistical analyses of real-world data\cite{Ning,Bayati}, and geospatial modeling\cite{Essel}. Indeed, the concern over unequal access to vaccines was the driving force behind the establishment of COVAX shortly after the onset of the pandemic\cite{Khairi,COVAX}. Through data-driven mathematical models, it has been estimated that more lives could have been saved and the global pandemic could have been curtailed sooner if equitable vaccine access had been achieved\cite{Ye,Moore,Gozzi,Li}. Factors contributing to this disparity include the initial shortage of sufficient vaccines\cite{Khairi}, intellectual property issues\cite{Pilkington}, and high-income countries underestimating the threat posed by new virus strains\cite{Ye}.\\

One of the primary drivers behind the unequal distribution of vaccines is the imperative for high-income countries to prioritize the vaccination of their own populations as soon as vaccines become available during the early stages of vaccination campaigns\cite{Puyvallee}. It is hard to persuade the nationals, and perhaps unfair to those of high income countries to share their vaccines, especially when the pandemic situations in their own countries are not under control. While establishing international initiatives like COVAX is essential for achieving equitable access, these global efforts can still be hindered by the self-interests of high-income nations. To address this issue and strive for greater equity in future pandemics, we advocate for the utilization of straightforward yet evidence-based mathematical models. These models can serve to persuade individuals in high-income countries that timely sharing of vaccines could effectively mitigate the spread of the pandemic within their own borders.\\ 

We modify the two coupled Susceptible-Infected-Recovered-Deceased (SIRD) model initially developed by J. Burton et. al.\cite{Burton} to illustrate the impact of vaccination and migration on the measles outbreak within specific subpopulations in Cameroon.  This model is adapted to examine the temporal progression of the COVID-19 pandemic in two hypothetical countries with potential migratory connections. In our study of COVID-19, two key model parameters, namely the infection rate and vaccination rate, are made time-dependent. The time-varying infection rate accounts for the impact of non-pharmaceutical interventions in curbing virus transmission and the emergence of pathogenic variants. Likewise, the time-varying vaccination rate reflects the commencement of large-scale vaccination efforts approximately one year after the pandemic's onset, with different rates employed to simulate scenarios involving vaccine sharing. The determination of the infection rate, initial vaccination rate magnitude, and other time-independent parameters within the model is achieved by comparing actual data sourced from the United States\cite{91COVID}, chosen for its reliability and comprehensiveness during the early stages of the pandemic, with numerical outcomes derived from a single-country SIRD model. Population exchange coefficients are determined based on international travel data from the United States during the COVID-19 crisis.\\

With parameters calibrated using real-world data, we examine the dynamics within two hypothetical countries. In this setup, Country 1 begins vaccinating its population 300 days after the pandemic's onset. Meanwhile, Country 2 initiates vaccination only after Country 1 starts sharing a predetermined vaccination rate. Across all scenarios, our findings indicate that Country 1 benefits from timely vaccine sharing to minimize its own pandemic duration, particularly when there's mutual population exchange between the two countries. Country 2 experiences the shortest pandemic duration if Country 1 shares its vaccines from the outset. The emergence of pathogenic virus variants influences the optimal timing for vaccine sharing in Country 1. Modeling these variants reveals that regardless of their presence, Country 1 still benefits from timely vaccine sharing to shorten its pandemic duration. These results align with previous findings from more complex mathematical models or statistical analyses\cite{Ye,Moore,Gozzi,Li,Ning,Bayati}, indicating that vaccine sharing is beneficial for Country 1. The key distinction in our work lies in the fact that while Country 1, as a high-income nation with early vaccine access, may not need to share vaccines immediately, doing so in a timely manner is advantageous for its own welfare.\\
      
The article is structured as follows: In Section \ref{ms}, we present the two coupled SIRD models with time-dependent model parameters. In Section \ref{modelp}, we detail the methodology for obtaining the time-dependent model parameters through comparison with real data. In Section \ref{twoc}, we analyze how the duration of the pandemic is affected by the distribution of vaccines for two hypothetical countries under different conditions. We also examine the influence of potential pathogenic virus variants in a separate subsection. In Section \ref{conclusion}, we provide a summary of our findings and discuss the potential implications of this research.

\section{Two coupled SIRD models}\label{ms}
To simulate the progression of the pandemic within two interconnected groups with initial populations N1 and N2, we utilize a modified SIRD model. This model incorporates vaccination to transition susceptible individuals to the recovered state, while also allowing for population flow between the two groups. The original equations referenced in Ref.~\onlinecite{Burton} employ model parameters $\beta_i$, $\kappa_i$, $\nu_i$, $\mu_i$, and an additional parameter related to birth rate, all assumed to be constant over time. We assume that the natural rates of birth and death, or what is termed vital dynamics, are insignificantly small compared to other parameters. Thus, we omit these two parameters in our model (death rate accounts for deaths due to COVID-19 only in this paper). The equations governing the number of susceptible $S_i(t)$, currently infected $I_i(t)$, recovered $R_i(t)$, and deceased $D_i(t)$ individuals within group $i$ ($i=1$ or $2$ representing distinct regions or countries) are expressed as follows:
\bea\nn
 \frac{dS_1(t)}{dt}&=&-\beta_1(t) \frac{S_1(t) (1-c_3)I_1(t)}{N_1} -\beta_1(t) \frac{S_1(t) c_3 I_2(t)}{N_1}-\nu_1(t) S_1(t)+c_1S_2(t)-c_2S_1(t),\\\nn
\frac{dI_1(t)}{dt}&=&\beta_1(t) \frac{S_1(t) (1-c_3)I_1(t)}{N_1}+\beta_1(t) \frac{S_1(t) c_3 I_2(t)}{N_1}-\kappa_1(t) I_1(t)
-\mu_1(t)I_1(t) +c_1I_2(t)-c_2I_1(t),\\\nn
 \frac{dR_1(t)}{dt}&=&\kappa_1(t) I_1(t) +\nu_1(t) S_1(t)+c_1R_2(t)-c_2R_1(t) ,\\\label{cSIRD model}
 \frac{dD_1(t)}{dt}&=&\mu_1(t)I_1(t).\\\nn
  \frac{dS_2(t)}{dt}&=&-\beta_2(t) \frac{S_2(t) (1-c_3)I_2(t)}{N_2} -\beta_2(t) \frac{S_2(t) c_3 I_1(t)}{N_2}-\nu_2(t) S_2(t)+c_2S_1(t)-c_1S_2(t),\\\nn
\frac{dI_2(t)}{dt}&=&\beta_2(t) \frac{S_2(t) (1-c_3)I_2(t)}{N_2}+\beta_2(t) \frac{S_2(t) c_3 I_1(t)}{N_2}-\kappa_2(t) I_2(t)
-\mu_2(t)I_2(t)+c_2I_1(t)-c_1I_2(t),\\\nn
 \frac{dR_2(t)}{dt}&=&\kappa_2(t) I_2(t) +\nu_2(t) S_2(t) +c_2R_1(t)-c_1R_2(t),\\\nn
 \frac{dD_2(t)}{dt}&=&\mu_2(t)I_2(t).
\eea

In this interconnected SIRD model, we assume the transmission of the disease occurs between infected individuals and susceptible individuals in a thoroughly mixed system within each region or country $i$. Any spatial disparities, such as variations in population density between urban and rural areas, are smoothed out within the model parameters $\beta_i(t)$, $\kappa_i(t)$, $\nu_i(t)$ and $\mu_i(t)$. These time-varying parameters in the epidemic compartment model have been discussed in various contexts\cite{Landry,Granger,Scoglio}. Here, $\beta_i(t)$ represents the time-varying infection rate for group $i$, which diminishes from its maximum value, determined by the nature of the virus and the average population density in the region under consideration, to a small positive value close to zero in scenarios where stringent non-pharmaceutical interventions (NPIs) are implemented. $\kappa_i(t)$ denotes the recovery rate, indicating the average duration for an infected individual to recover from the disease. $\nu_i(t)$ signifies the vaccination rate aimed at transitioning susceptible individuals to the recovered state, while $\mu_i(t)$ represents a time-varying mortality rate. The parameters $c_1$, $c_2$, and $c_3$ 
respectively denote the coefficients for population exchange from group two to group one, from group one to group two, and for interactions between infected individuals and susceptible individuals in the other group. These parameters concerning population movement between the two groups are assumed to be constant over time, although adjustments for time-varying parameters could be made straightforwardly if real-time travel data between the groups were available.\\

In the following discussion, we limit the consideration of time-dependent parameters to the infection rate $\beta_i(t)$ and vaccination rate $\nu_i(t)$ exclusively. While theoretically, all parameters could vary over time, the accurate estimation and interpretation of these parameters heavily rely on the availability and reliability of real-world data. Furthermore, introducing time-dependent parameters in the SIRD model complicates the stability analysis of the associated differential equations\cite{Burton,Granger}, often necessitating numerical solutions. To mitigate this complexity, we opt to make only  $\beta_i(t)$ and $\nu_i(t)$ time-dependent, simplifying the analysis. The parameter $\beta_i(t)$ reflects the ease of virus transmission, which can be influenced by non-pharmaceutical interventions like mask mandates and social distancing measures. Additionally, the emergence of COVID variants may temporarily increase $\beta_i(t)$. On the other hand, $\nu_i(t)$ represents the timing and distribution of vaccines during the pandemic. Both factors significantly impact the trajectory of the pandemic, as illustrated in the subsequent section using a single-country scenario as an illustration. 

\section{Method to obtain model parameters}\label{modelp}
For a single country scenario, Eq.(\ref{cSIRD model}) simplifies to:
\bea\nn
 \frac{dS(t)}{dt}&=&-\beta(t) \frac{S(t) I(t)}{N} -\nu(t) S(t),\\\nn
\frac{dI(t)}{dt}&=&\beta(t) \frac{S(t) I(t)}{N}-\kappa I(t)
-\mu I(t) ,\\\nn
 \frac{dR(t)}{dt}&=&\kappa I(t) +\nu(t) S(t) ,\\\label{SIRD model}
 \frac{dD(t)}{dt}&=&\mu I(t).
\eea 
In Eq.(\ref{SIRD model}), the subscripts denoting different regions or countries are removed, and the parameters $c_i$ related to population exchange between different regions/countries are also omitted. In this simplified context, vaccination directly transitions susceptible individuals to the recovered state, and once individuals are in the recovered category, they do not revert back to the susceptible population. This model does not account for reinfection with COVID-19 or the decline in vaccine efficacy against infection over time, as reported in recent studies\cite{Feikin}.\\ 

The reinfection rate, averaging around 5$\%$ according to several studies\cite{Guedes,Flacco,Medic}, varies notably across different demographic groups such as age, household incomes, and vaccinated or unvaccinated status. However, it exhibits less variability based on the number of vaccine doses received per person\cite{Medic}. Incorporating reinfection into the model extends the time required to achieve a disease-free state\cite{Burton} but does not significantly prolong the duration to end the pandemic, defined here as the point where no Non-Pharmaceutical Interventions (NPIs) are deemed necessary in that region. The latter definition is widely accepted as the endpoint of the pandemic\cite{Ioannidis}. We adhere to this definition and omit the reinfection issue from the subsequent discussion.\\

We utilize real data on the cumulative number of infections over time in the United States (USA), sourced from Ref.~\onlinecite{91COVID}, a data visualization website developed by Prof. Fagen-Ulmschneider at the University of Illinois. This dataset spans from April 2020 to March 2022, covering approximately 700 days. Employing this authentic data and following the methodology outlined in Ref.~\onlinecite{Chao}, we get the infection rate $\beta(t)$, vaccination rate $\nu(t)$, recovery rate $\kappa$, and death rate $\mu$ as: $\beta(t)=\ln(\beta_0(t))+0.0025$
, $\nu(t)=10^{-2}\Theta(t-300)=\nu_0\Theta(t-300)$, $\kappa=5.5\times 10^{-3}$, $\mu=1.8\times 10^{-4}$. In this context, the unit of time $t$, is in "days." Vaccination commences approximately 300 days into the observation period. Here, $\beta_0(t)$ represents an empirical function fitted to the actual data, and its explicit form is provided below:

\bea\nn
\beta_0(t)&=&\theta (t-14) \Large[5.52267 \exp \left(-0.743924 \sqrt{t}\right)+\frac{1}{900} 0.03 (t-630) (690-t) \theta (690-t) \theta (t-630)\\\nn
&+&\frac{1}{400} 0.01 (t-530) (570-t) \theta (570-t) \theta (t-530)+\frac{1}{330} 0.02 (457-t) \theta (457-t) \theta (t-127)\\\nn
&+&\frac{1}{32} 0.02 (t-95) \theta (127-t) \theta (t-95)+1.00022\Large]+1.35 \theta (14-t);
\eea
   
\begin{figure}
\centering
\begin{subfigure}{0.48\textwidth}
\includegraphics[width=\textwidth]{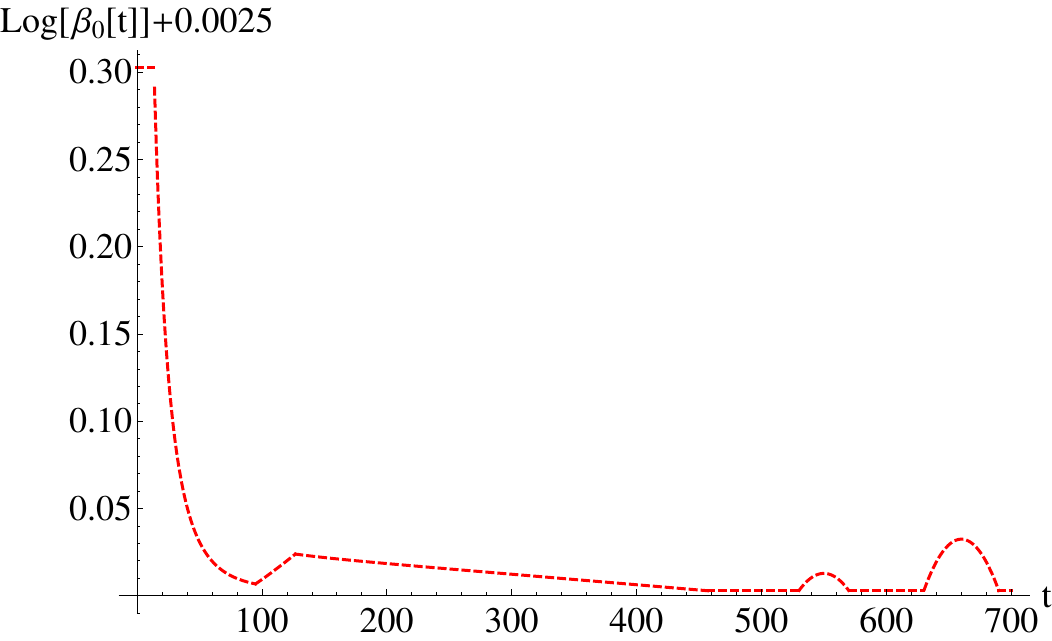}
\caption{Infection rate versus time.}
\label{fig:betausa1}
\end{subfigure}
\hfill
\begin{subfigure}{0.48\textwidth}
\includegraphics[width=\textwidth]{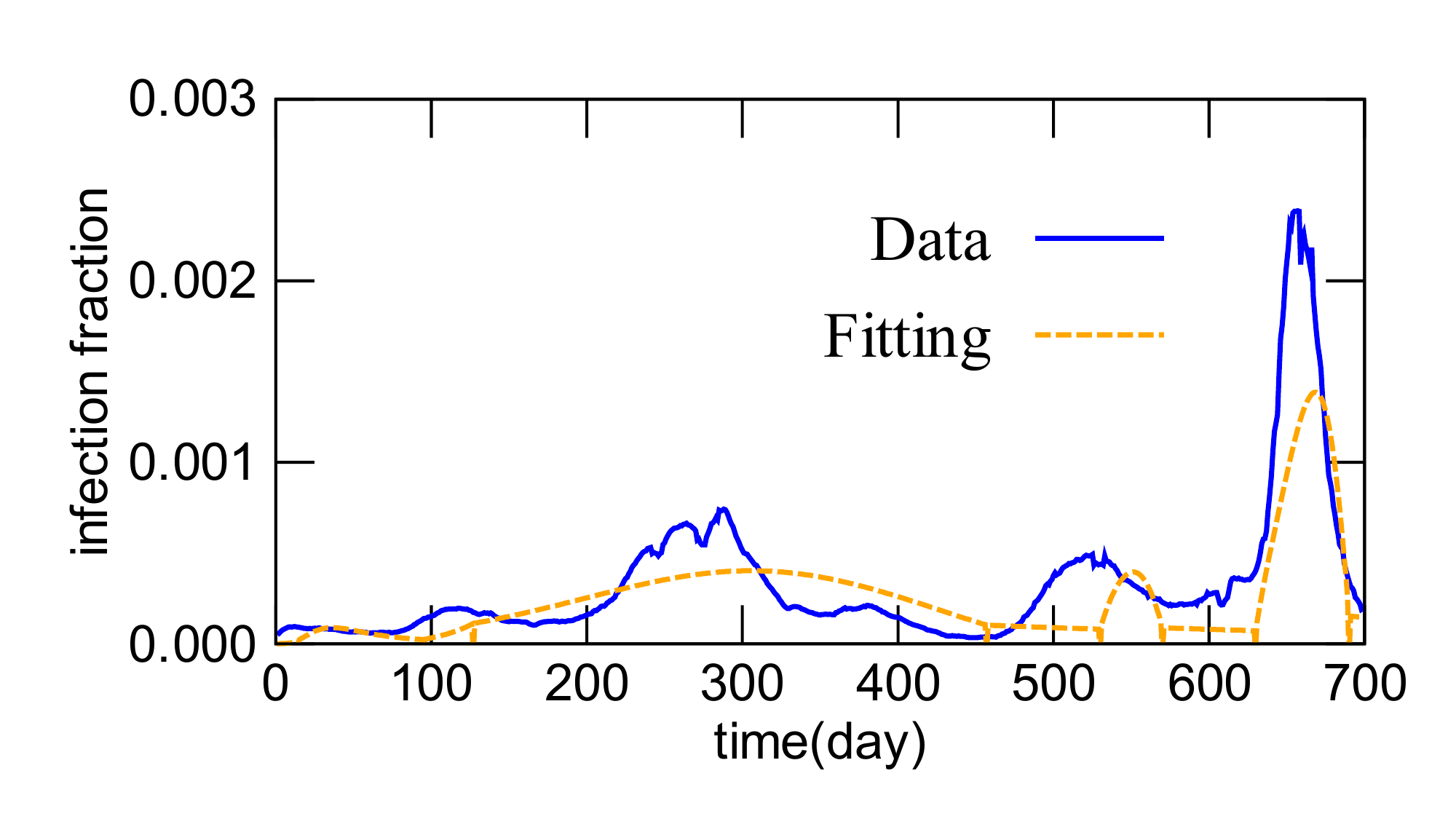}
\caption{Daily infection fraction versus time}
\label{fig:betausa2}
\end{subfigure}
\caption{(a) Infection rate $\beta(t)$ versus time $t$. Time is measured in days. (b) Contrasting the daily fraction of confirmed infections (averaged over seven days) from April 2020 to March 2022, the blue line depicts actual data, while the orange dashed line illustrates simulated data using the infection rate from Fig.\ref{fig:betausa1} for the USA. The total population is normalized to one, and time is measured in days.}
\end{figure}

\begin{figure}
\centering
\begin{subfigure}{0.48\textwidth}
\includegraphics[width=\textwidth]{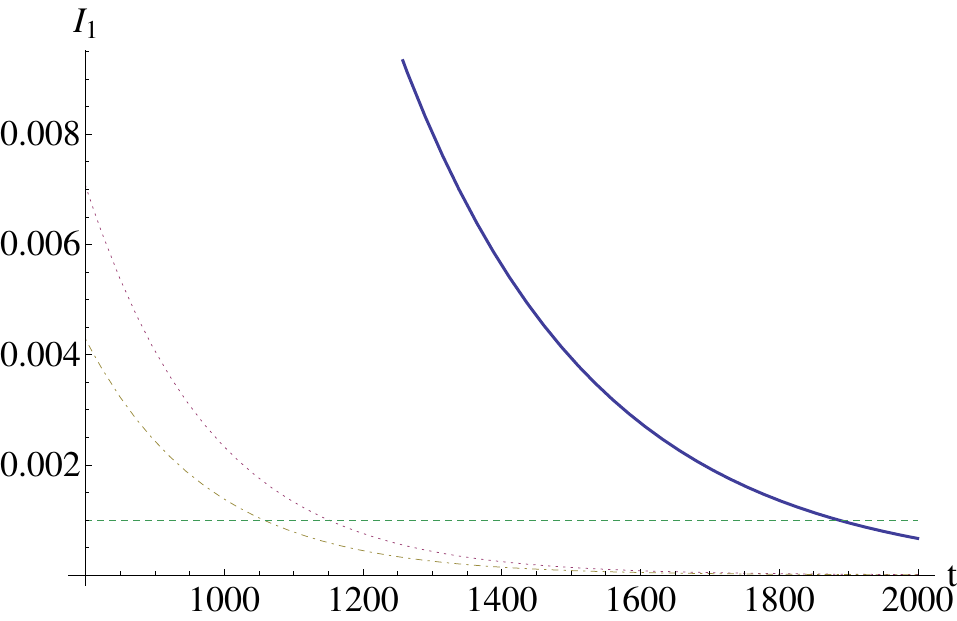}
\caption{Infected population v.s. time.}
\label{fig:tausa1}
\end{subfigure}
\hfill
\begin{subfigure}{0.48\textwidth}
\includegraphics[width=\textwidth]{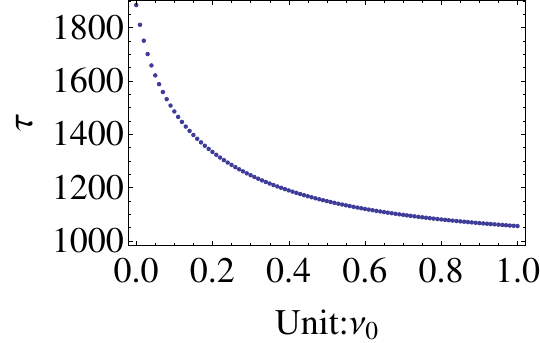}
\caption{Main:  Inset: $\tau$ v.s. $\nu$.}
\label{fig:tausa2}
\end{subfigure}
\caption{(a) The plot illustrates the infected fraction over time for three scenarios: when $\nu=0$  (blue line), $\nu=0.5\nu_0$  (purple dotted line), and $\nu=\nu_0$ (brown dot-dashed line) nearing the end of the pandemic. Additionally, the green dashed line marks the condition signifying the end of the pandemic. (b) Epidemic control time $\tau$ versus various magnitudes of vaccination rates, denoted by $\nu$, ranging from zero to $\nu_0$ in units of $\nu_0$}
\end{figure}

The function $\beta(t)$ plotted against time is depicted in Fig.\ref{fig:betausa1}. Fig.\ref{fig:betausa2} shows the actual data (seven days average of daily new infected fraction from Ref.~\onlinecite{91COVID}, blue line) compared with the one generated by the function $\beta(t)$ (orange dashed line). The fitting could be improved if more sophisticated choice of functional form is chosen. For example, the sharp changes in orange line at day 95, 127, 457, etc. are related to the choice of step functions in time in the $\beta_0(t)$ and could be smoothened if a continuous first derivatives in $\beta_0(t)$ were chosen. The choice of fitting function is not unique and subject to the change in other parameters, such as recovery and death rate, in the SIRD model. For example, the height of the last peak in blue line could be increased if maximum of quadratic function and the recovery rate in that region were increased.\\

From the function $\beta(t)$ describing the pandemic in the U.S.A., we observe commonalities shared with other nations. For instance, the initial peak marking the onset of the pandemic tends to be the highest regardless of whether subsequent peaks stem from domestic or foreign virus variants. While the intervals, durations, and heights of successive peaks lack a consistent pattern, they consistently remain lower than the first peak. In Fig.\ref{fig:betausa1}, the three subsequent peaks in the U.S.A. roughly align with COVID-19 variants designated by the World Health Organization (W.H.O.) as alpha, delta, and omicron. These virus variants all originated elsewhere and can be regarded as imports\cite{wikiv}. In contrast, countries like South Africa, India, and the United Kingdom have experienced both domestically originating and imported virus variants, yet the shape of the peaks in the function $\beta(t)$ cannot definitively distinguish between native and foreign variants.\\

In the United States, the large-scale vaccination campaign begins approximately 300 days after the initial outbreak, as previously mentioned. The vaccination rate, denoted by $\nu(t)=10^{-2}\Theta(t-300)$, is extrapolated based on the administration of $3.5$ million vaccine doses per day on average during the peak period of vaccination, spanning from $t=300$ to $t=700$\cite{91COVID,footnote1}. We analyze the time required to achieve epidemic control, herein referred to as the "epidemic control time," by comparing it against various hypothetical vaccination rates. It's important to note that "the end of the pandemic" in this context does not signify complete disease eradication, as mentioned in Ref.~\onlinecite{Burton}, but rather is defined as the point beyond which the proportion of infected individuals consistently remains below a certain threshold, related to the country's medical capacity. In this article, this threshold is set at $10^{-3}$. The choice of this threshold value is informed by the number of hospital beds per thousand people, which is $2.8$ in the U.S.A.\cite{UShos}. This threshold is depicted as a green dashed line in Fig.\ref{fig:tausa1}.\\

The main takeaway is that the infection rate $\beta(t)$ observed in other countries exhibits a similar pattern and characteristics\cite{Chao} as demonstrated in Fig.\ref{fig:betausa1} for the U.S.A., albeit with different functional forms and durations of sub-peaks. Subsequently, we apply the model parameters derived from the U.S.A., with certain modifications outlined in the following section, to two hypothetical countries. Country 1 initiates its vaccination campaign on day 300, while the commencement day and vaccination rate for Country 2 are determined by Country 1, mirroring a scenario where Country 1 shares its vaccine resources with Country 2.      
   
\section{Cases of two countries}\label{twoc}
We utilize the mutual migration rate $c_i$ in Eq.\ref{cSIRD model} to represent the frequency of population exchanges between the two countries. For two identical countries, the mutual migration rate is set as $c_1=c_2=1.42\times 10^{-5}$, and $c_3=7.0\times 10^{-7}\simeq 0.05c_1$. When dealing with countries of differing population ratios, the migration rate for Country 2 is adjusted as $c_2=c_1\times (N_2/N_1)$ to maintain a fixed total population in the absence of fatalities. The value
$c_1=c_2=1.42\times 10^{-5}$ is estimated based on inbound arrivals to the USA, which totaled around 19.2 million in 2020\cite{sta,ft1}. This figure represents a decrease compared to other years due to travel restrictions imposed during the pandemic. It's worth noting that migration rates $c_i$ vary significantly among different countries.

\subsection{Comparisons for four scenarios}\label{foursce}
Here, we examine how various factors such as different population sizes $N_i$, infection rate $\beta_i(t)$, and timing of vaccine distribution impact the duration of the pandemic in two countries. For infection rate $\beta_i(t)$, we utilize the infection rate $\beta_0(t)$ for the U.S.A. case and extend it from day 690 to remain constant until the end of the pandemic. These conditions are categorized into four scenarios, as outlined in Table \ref{tab:4cases}. Within each scenario, we further discuss cases with and without mutual migrations between the two countries.\\

\begin{table}[h]
\begin{tabular}{|l|l|l|}
\hline
1. $\beta_1=\beta_2$, $N_1=N_2$, $c_i=0$ or $c_i\neq 0$&2. $\beta_1\neq\beta_2$, $N_1=N_2$, $c_i=0$ or $c_i\neq 0$\\
\hline
3. $\beta_1=\beta_2$, $4N_1=N_2$, $c_i=0$ or $c_i\neq 0$&4. $\beta_1\neq\beta_2$, $4N_1=N_2$, $c_i=0$ or $c_i\neq 0$\\
\hline
\end{tabular}
\caption{Four scenarios with different combinations of population ratio $N_1/N_2$ and infection rate $\beta_i(t)$.}
\label{tab:4cases}
\end{table}

In the first scenario, both countries share identical populations and infection rates. Moving to the second scenario, the infection rate of country 2 experiences a 30-day delay. This delay is represented by $\beta_2(t)=\beta_1(t-30)$ for $t>30$ and $\beta_2(t)=0$ for $0<t\le 30$, effectively postponing the onset of the pandemic in country 2 by 30 days. The third and fourth scenarios mirror the first and second ones, respectively, with the key distinctions being the population ratio $N_2/N_1$ and mutual migration rate $c_2/c_1$ is changed from $1$ to $4$. These adjustments illustrate how a larger susceptible population alters the trajectory of the pandemic.\\

\begin{figure}
\centering
\begin{subfigure}{0.48\textwidth}
\includegraphics[width=\textwidth]{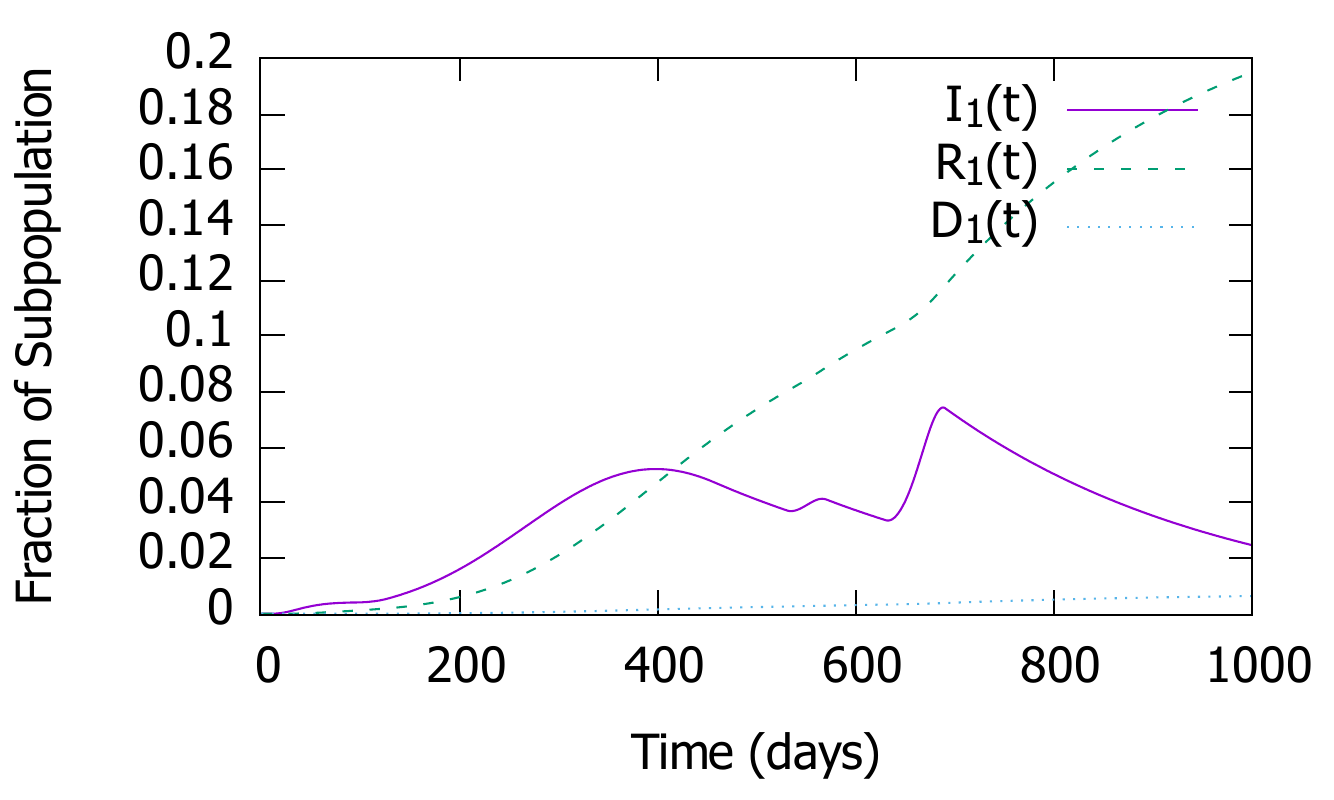}
\end{subfigure}
\hfill
\begin{subfigure}{0.48\textwidth}
\centering
\includegraphics[width=\textwidth]{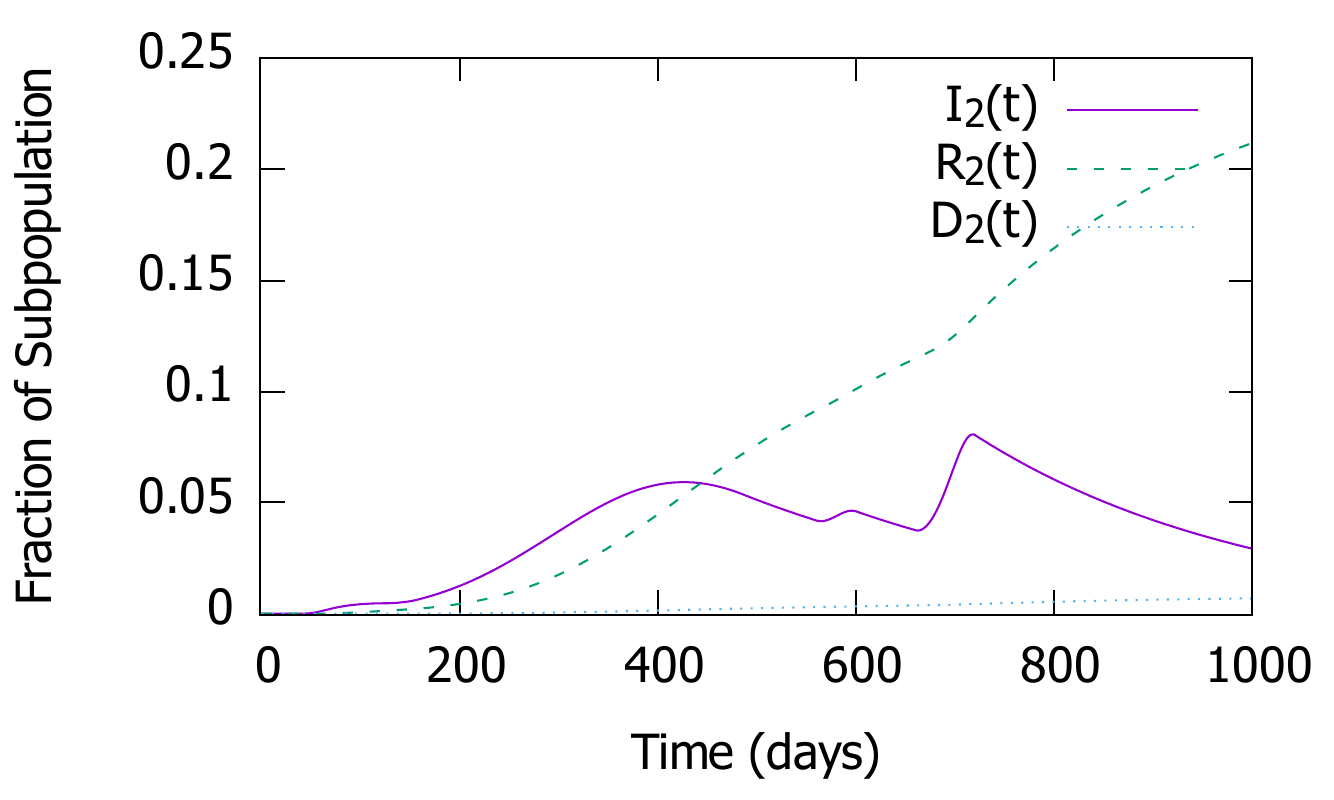}
\end{subfigure}
\caption{The Infected-Recovered-Deceased population ratio for the country 1 (left) and country 2(right) as a function of time in scenario 2 with $c_i\neq0$. Vaccination rate $\nu_1=\nu_2=0$ here.}
\label{fig:dird1ird2(commute)}
\end{figure}

The reasons for implementing different starting times in scenarios 2 and 4 across various countries during the pandemic serve dual purposes. Firstly, the initial outbreak typically pinpoints the virus's origin, and simultaneous outbreaks in non-bordering countries are uncommon. Secondly, it illustrates how mutual migrations between two countries alter the initial conditions and subsequent course of the pandemic. In instances where there are no mutual migrations ($c_i=0$) between two countries, the infected ratios, $I_1(t)$ and $I_2(t)$, exhibit identical patterns, albeit with a time shift. However, when mutual migrations are taken into account ($c_i\neq 0$), it becomes apparent that the epidemic in country 2 surpasses that in country 1. This is evident from the fact that the peak value of $I_2(t)$ exceeds that of $I_1(t)$ as shown in Fig.\ref{fig:dird1ird2(commute)}. The rationale behind this is that country 2 accumulates a greater number of infected individuals at
$t=30$ due to mutual migrations, resulting in a higher infected ratio despite having the same infection rate magnitude.\\

\begin{figure}
\centering
\begin{subfigure}{0.48\textwidth}
\includegraphics[width=\textwidth]{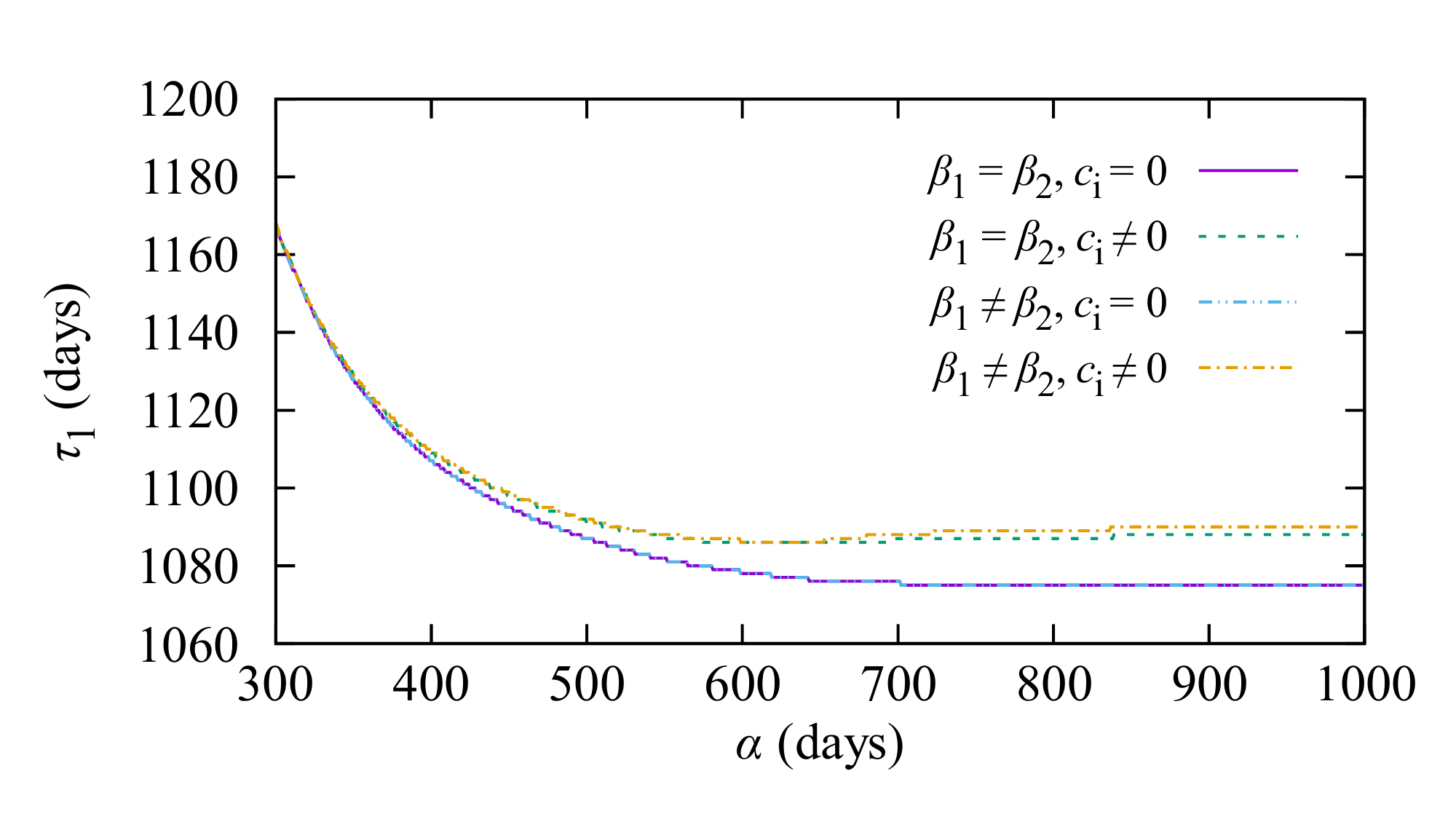}
\caption{Scenario 1 and 2 for country 1}
\label{fig:tau1}
\end{subfigure}
\hfill
\begin{subfigure}{0.48\textwidth}
\includegraphics[width=\textwidth]{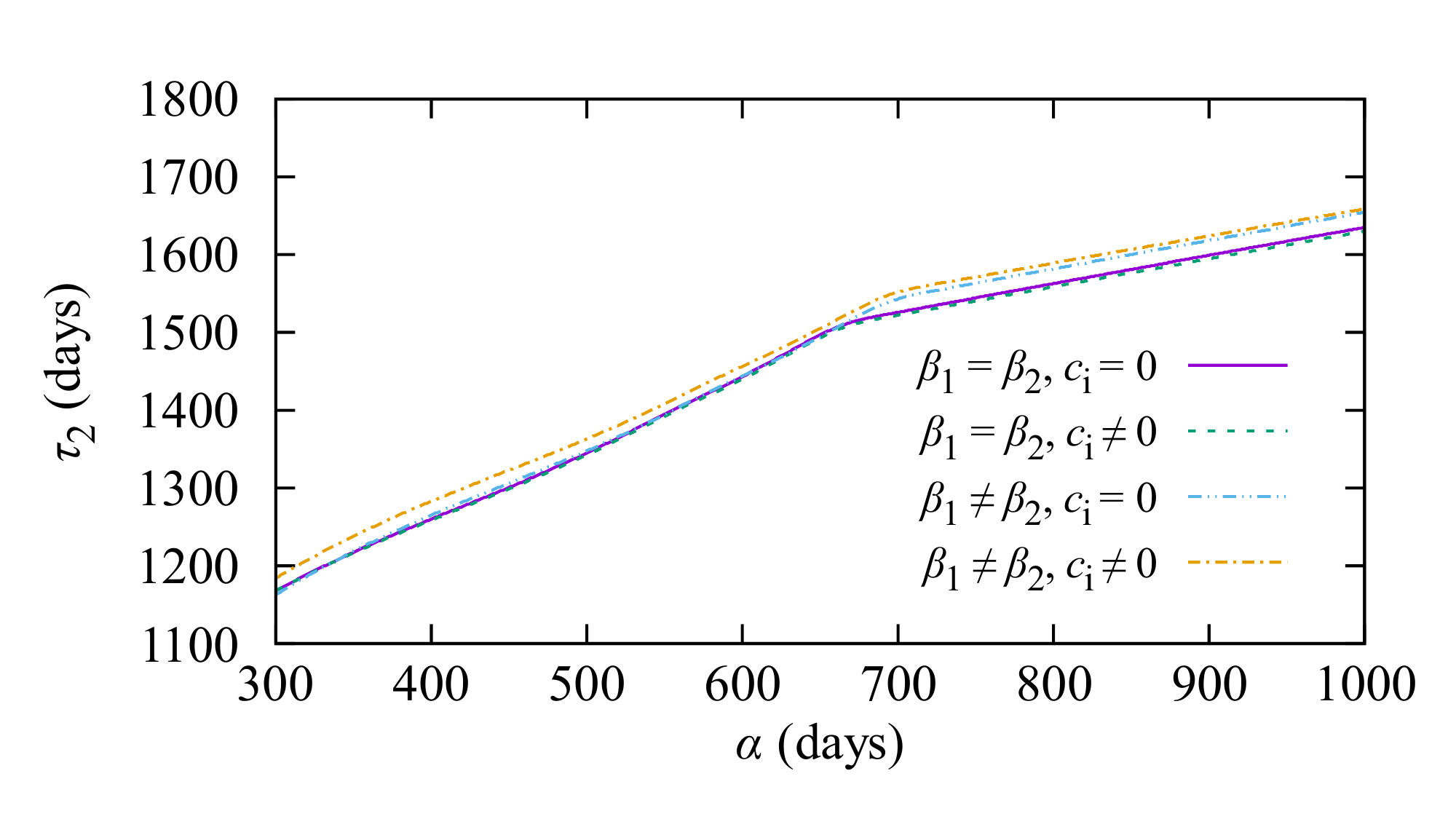}
\caption{Scenario 1 and 2 for country 2}
\label{fig:tau2}
\end{subfigure}
\hfill
\begin{subfigure}{0.48\textwidth}
\includegraphics[width=\textwidth]{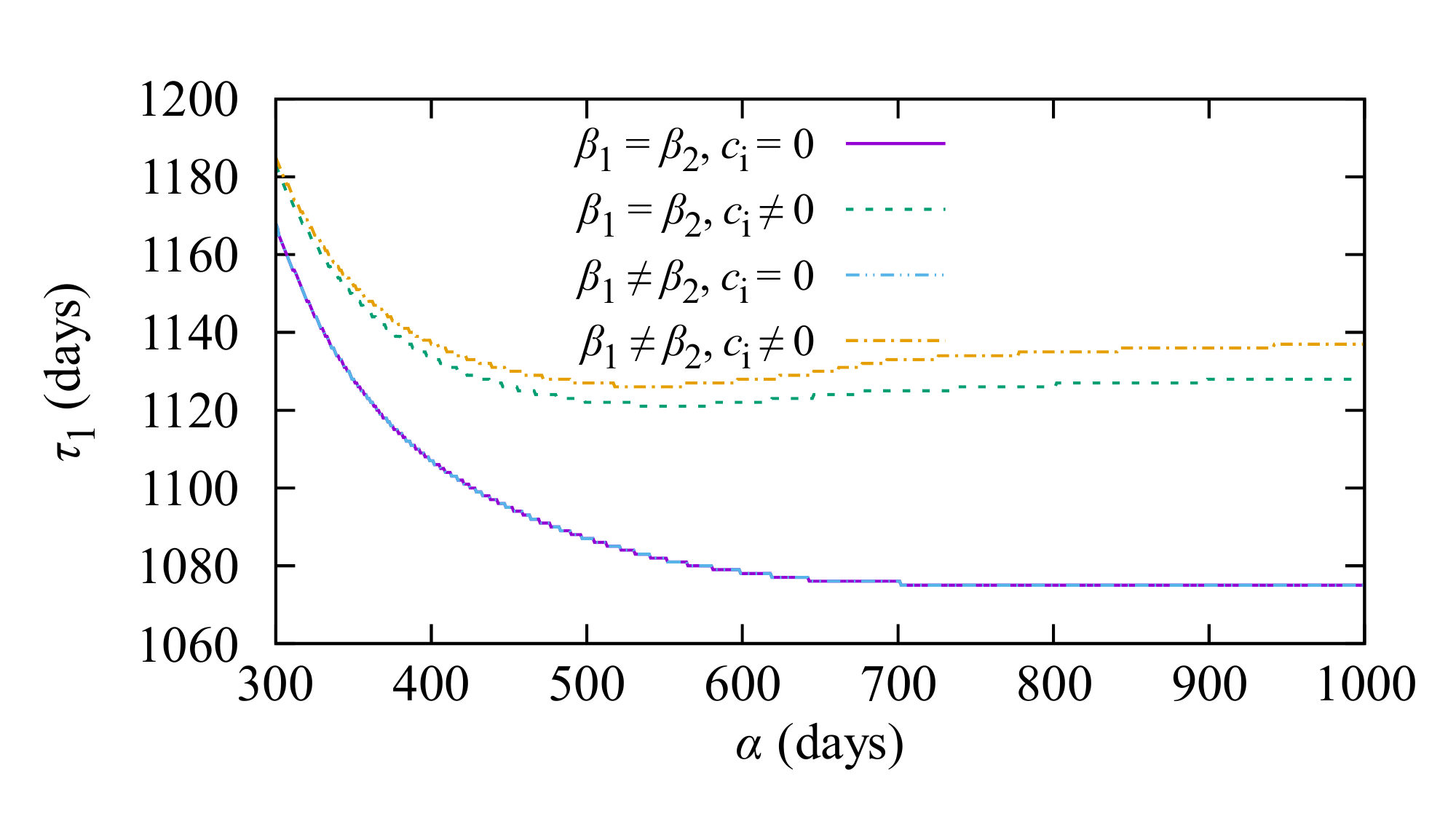}
\caption{Scenario 3 and 4 for country 1}
\label{fig:4tau1}
\end{subfigure}
\hfill
\begin{subfigure}{0.48\textwidth}
\includegraphics[width=\textwidth]{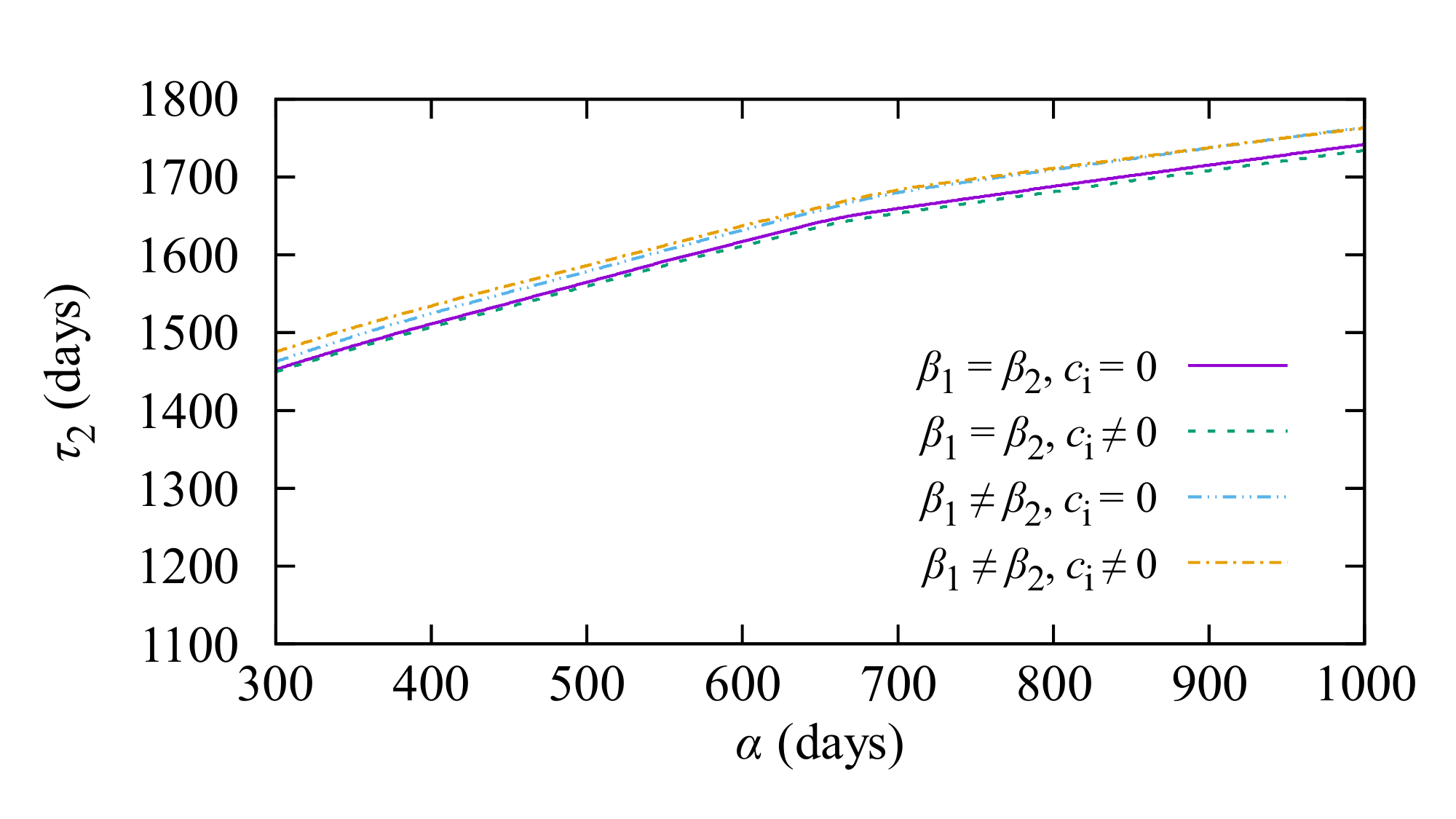}
\caption{Scenario 3 and 4 for country 2}
\label{fig:4tau2}
\end{subfigure}
\caption{Epidemic control time $\tau_i$ for country $i$ ($i=1,2$) as a function of vaccine distribution time $\alpha$ for the four scenarios listed in the Table.\ref{tab:4cases}. The purple line and the cyan dotted line ($c_i=0$ cases) completely overlap in both Fig.\ref{fig:tau1} and Fig.\ref{fig:4tau1}. The overlap is caused by the absence of interaction between the two countries.}
\label{fig:tau1tau24tua14tau2}
\end{figure}

On the 300th day, country 1 commences vaccinating its populace against COVID-19. Subsequently, country 1 determines when to distribute a portion of the vaccines to country 2, which lacks its own vaccine production capabilities. This distribution entails reducing the vaccination rate $\nu_1$ to half of its original value, while increasing country 2's vaccination rate $\nu_2$ from zero to half of country 1's original vaccination rate. We designate as the day when country 1 begins sharing the vaccines. After this day, both vaccination rates, $\nu_1$ and $\nu_2$, remain constant over time. We denote $\tau_1$ and $\tau_2$ as the epidemic control times for country 1 and country 2, respectively.\\

The graph depicts the epidemic control time as a function of vaccine allocation time in both country 1 and country 2. In scenarios 1 and 2, illustrated in Fig.\ref{fig:tau1} and Fig.\ref{fig:tau2}, and scenarios 3 and 4, shown in Fig.\ref{fig:4tau1} and Fig.\ref{fig:4tau2}, respectively, the trend is examined. From the purple line in both Fig.\ref{fig:tau1} and Fig.\ref{fig:tau2}, we see that if the country 1 has immediately allocated vaccines since its acquisition of the vaccines on the 300th day, the epidemic control times of both countries are the same, i.e. on the 1168th day for the first scenario without mutual migrations. However, if country 1 starts to distribute vaccine resources after the 300th day, the epidemic control time of country 1 would be shortened as expected, whereas the control time in country 2 would be significantly prolonged. \\

Let's begin by analyzing the epidemic control time concerning vaccine allocation time in country 1. Observing Fig.\ref{fig:tau1} and Fig.\ref{fig:4tau1}, it's apparent that in the absence of mutual migrations, the epidemic control time in country 1 $\tau_1$ steadily decreases as the vaccine allocation time $\alpha$ increases. This trend arises because individuals infected in country 2 cannot transmit COVID through mutual migrations. Furthermore, due to $c_i=0$, the purple line and the cyan dotted line completely overlap in both Fig.\ref{fig:tau1} and Fig.\ref{fig:4tau1}.\\

Conversely, when $c_i\neq 0$, we observe that the epidemic control time $\tau_1$ exhibits a global minimum. This indicates that country 1 has an optimal timing for vaccine allocation. Relative to $c_i=0$, mutual migrations between the two countries lead to a delay in epidemic control time. Therefore, a stringent border control strategy, which reduces $c_i$, proves more advantageous for country 1 in pandemic control efforts. Similarly for $N_2=4N_1$, the end of the pandemic in country 1 $\tau_1$ becomes larger due to mutual migrations. As seen from the Fig.\ref{fig:4tau1} and Fig.\ref{fig:tau1}, the $\tau_1$ for $N_2=4N_1$ is larger than $\tau_1$ for $N_2=N_1$ at the same vaccine allocation time $\alpha$.\\

In the cases of country 2 depicted in Fig.\ref{fig:tau2} and Fig.\ref{fig:4tau2}, the epidemic control time $\tau_2$ exhibits a consistent increase as the vaccine allocation time $\alpha$ rises. This suggests that the sooner country 2 receives vaccines, the faster the pandemic in country 2 can be brought under control. Notably, the slope in both Fig.\ref{fig:tau2} and Fig.\ref{fig:4tau2} experiences a sudden decrease around the 700th day. This occurrence is attributed to the infection rate remaining unchanged after either day 690 or day 720, maintaining a small, constant value thereafter.\\ 

The population size in country 2 emerges as the pivotal factor determining the epidemic control day $\tau_2$. Upon comparison between Fig.\ref{fig:tau2} and Fig.\ref{fig:4tau2}, it becomes evident that the disparity in epidemic control time $\tau_2$ spans from approximately 100 to 250 days across various vaccine allocation times $\alpha$ for different population ratios $N_2/N_1$. Conversely, the difference $\tau_2$ at a fixed $N_2/N_1$ (represented by the four lines within Fig.\ref{fig:tau2} or Fig.\ref{fig:4tau2}) remains less than 50 days for varying infection rates or migration rates. This phenomenon can be attributed to the relatively small empirical value of the migration rate utilized and the slight 30-day shift in infection rates. Consequently, this does not induce a sharp change in the number of infected cases.\\

\begin{figure}
\centering
\begin{subfigure}{0.48\textwidth}
\includegraphics[width=\textwidth]{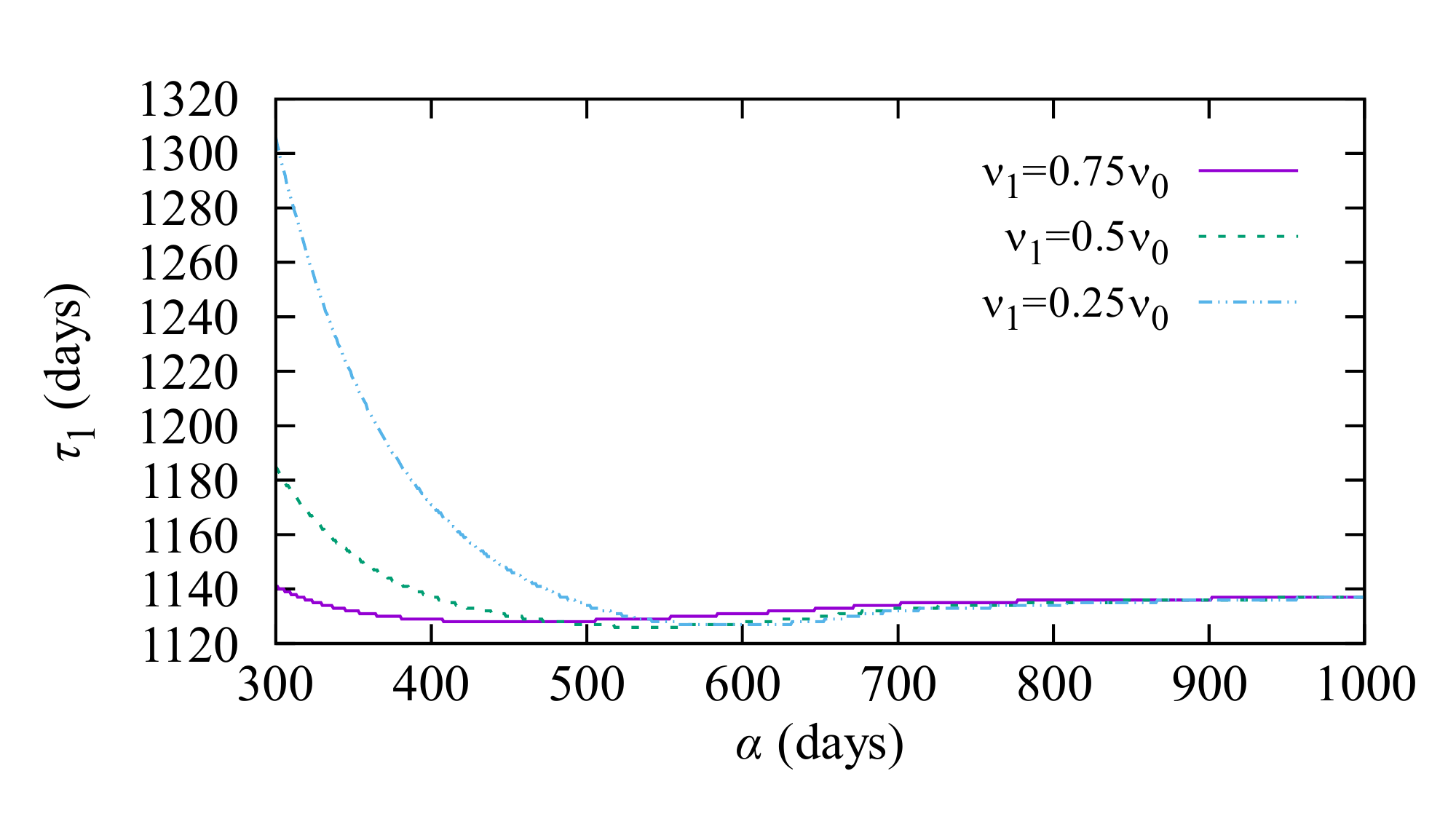}
\caption{Different vaccine allocation strategies in country 1}
\label{fig:dv1}
\end{subfigure}
\hfill
\begin{subfigure}{0.48\textwidth}
\includegraphics[width=\textwidth]{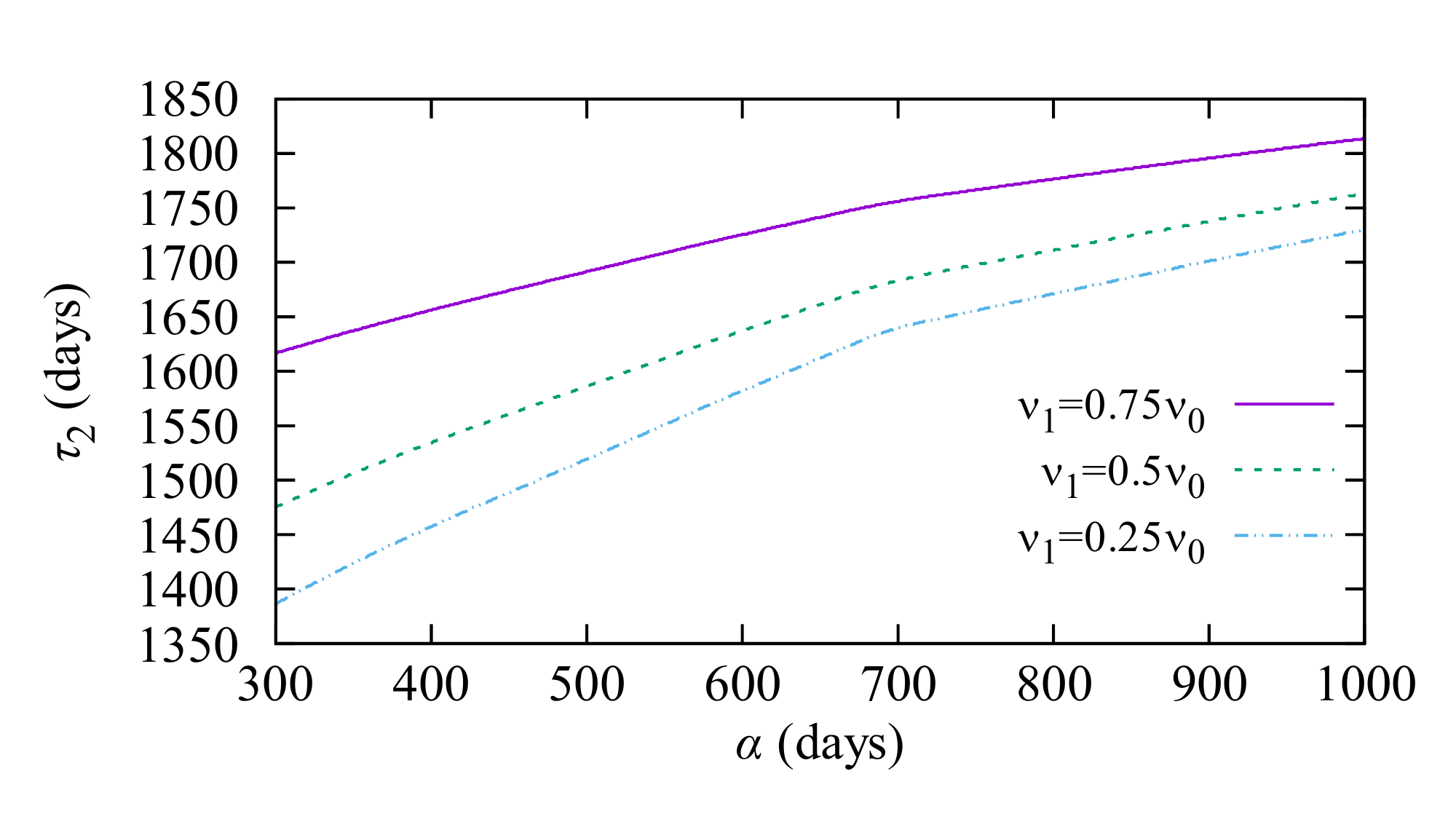}
\caption{Different vaccine allocation strategies in country 2}
\label{fig:dv2-1}
\end{subfigure}
\caption{Epidemic control time $\tau_i$ for country $i$ ($i=1,2$) as a function of vaccine distribution time $\alpha$ for three different amount of shared vaccines: $\nu_1=0.75\nu_0$, $\nu_1=0.5\nu_0$, and $\nu_1=0.25\nu_0$. Cases shown here are belong to scenario 4 with population exchange. Larger $\nu_1$ means country 1 keeps more vaccines and shares less.}
\label{fig:dv1v2}
\end{figure}

The quantity of vaccines shared by country 1 also significantly impacts the trajectory of the pandemic in both countries. Taking the case where $\beta_1\neq\beta_2$, $4N_1=N_2$, $c_i\neq 0$, or the fourth scenario involving population exchange as an example, we observe that the epidemic control time for country 1 $\tau_1$ exhibits a local minimum irrespective of the amount of vaccines shared by country 1, as depicted in Fig.\ref{fig:dv1}. When country 1 allocates more vaccine resources, such as sharing $3\nu_0/4$  of its vaccines with country 2 (equivalent to $\nu_1=0.25\nu_0$, represented by the cyan dotted line in Fig.\ref{fig:dv1}), the minimum of $\tau_1$ is reached at a larger vaccine allocation time $\alpha$. This indicates that fewer vaccines render it more challenging for country 1 to control its own epidemic. This altruistic act not only yields apparent benefits for country 2, as illustrated in Fig.\ref{fig:dv2-1}, but also minor benefits for country 1 at larger vaccine allocation times (for $\alpha>600$). This is attributed to mutual population exchange, which increases the infected population in country 1 if the pandemic in country 2 is not well controlled.\\

After $\alpha>900$, the differences between the three different $\nu_1$ become negligible in Fig.\ref{fig:dv1} but remain almost the same in Fig.\ref{fig:dv2-1}. Combining these two figures, we observe that if the vaccines were distributed at the minimum of $\tau_1$ in Fig.\ref{fig:dv1} (the minimum of $\tau_1$ for the three different $\nu_1$ are all around $1130$ but located at different $\alpha$ values), $\tau_2$ decreases with decreasing $\nu_1$ in Fig.\ref{fig:dv2-1}. The difference in $\tau_2$ between $\nu_1=0.5\nu_0$ and $\nu_1=0.25\nu_0$ is small compared to that between $\nu_1=0.75\nu_0$ and $\nu_1=0.5\nu_0$. This indicates that the effect of providing more vaccines at a later time is roughly equivalent to that of providing fewer vaccines at an earlier time.\\

In summary, we observe how the vaccine distribution time $\alpha$ influences the epidemic control time in both countries. Without migrations ($c_i=0$), country 1 can reduce its epidemic control time $\tau_1$ by approximately 100 days (refer to Fig.\ref{fig:tau1} or Fig.\ref{fig:4tau1}) if it delays the vaccine allocation time $\alpha$ until $t\simeq 700$. However, this strategy results in an extension of around 200 days (see Fig.\ref{fig:4tau2}) or 300 days (see Fig.\ref{fig:tau2}) for country 2. Additionally, sharing vaccines after around $t\simeq 700$ shows no noticeable decrease in $\tau_1$ but a significant increase in $\tau_2$. In essence, the epidemic control time in country 1 approaches a constant, while in country 2, it increases almost linearly with large values of $\alpha$. With mutual migrations, the epidemic control time in country 1 slightly increases after $\alpha$ surpasses 500 (see Fig.\ref{fig:4tau1}) or 600 (see Fig.\ref{fig:tau1}). Consequently, postponing vaccine sharing offers some benefits to country 1 initially but ceases to provide additional benefits after a certain point. Conversely, if country 2 receives vaccine supplies after $t\simeq 700$, $\tau_2$ grows almost linearly with $\alpha$. The more vaccines shared by country 1, the later (larger $\alpha$) it may start sharing to achieve minimal $\tau_1$ or to attain the same $\tau_2$. In essence, country 1 is advised to allocate vaccine resources no later than a certain time for its own benefit.\\

Next, we show that considering the possibility of new waves of infection due to virus variants, it is prudent to share vaccines with other countries to halt another wave of the pandemic.

\subsection{New waves of COVID from virus variants}
The emergence of COVID variants has extended the duration of the pandemic and also led to vaccines being less effective against the virus. In compartmental models like the SIRD model employed in this study, the introduction of a new pathogenic virus variant results in an increase in the infection rate $\beta(t)$, as illustrated in Fig.\ref{fig:betausa1} for the U.S.A. case. The resurgence of $\beta(t)$ can stem from various factors, but our focus here is on issues associated with pathogenic virus variants.\\

The conditions for the emergence of virus variants, whether originating from within its own country or imported from other countries, are expected to be distinct. Different infection rates are imposed to help distinguish theses two different sources of virus variant. For the virus variants generated in its own country, the infected population ratio $I(t)$ has to be higher than $I_m=10^{-2}$ for more than two weeks. This means that we need both sufficient number of infected population and sufficient long duration to cultivate the domestic pathogenic virus variant. For the rise of infection rate due to virus variant brought in from other country to happen, the infected population ratio $I(t)$ of the affected country has to go up by $10^{-6}$ irrespective of the duration. The increment of $I(t)$ is the same as the initial infected population ratio in the very beginning of the COVID pandemic in this study. It's important to note that the pandemic durations of domestic and imported virus variants are set to be mutually exclusive in this setup. While this may not fully reflect reality, it aids in clearly distinguishing the time sequence of domestic and imported virus variants. \\

\begin{figure}[h]
\centering
\begin{subfigure}{0.48\textwidth}
\includegraphics[width=\textwidth]{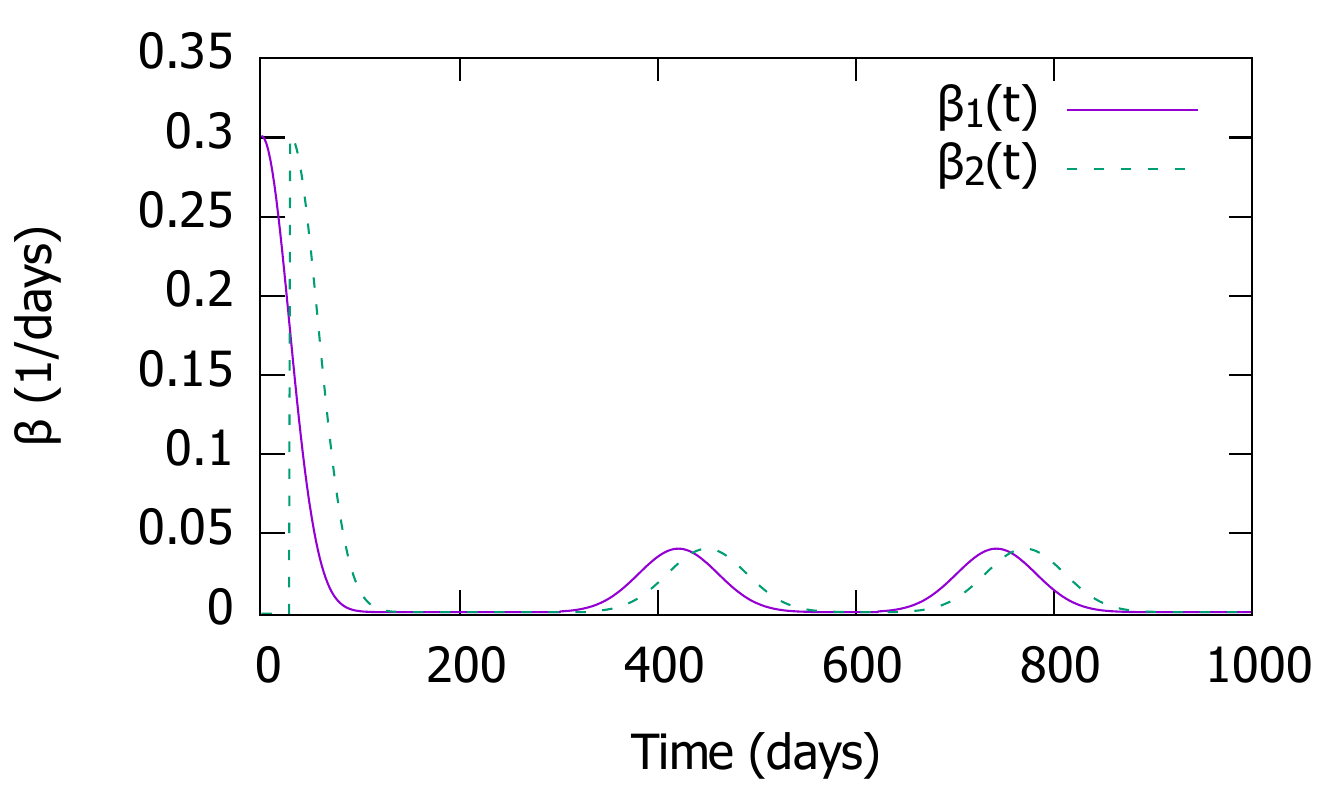}
\caption{The infection rate v.s. time with vaccine allocated on day 300.}
\label{fig:beta(na0.04ex0.06c10alpha300)}
\end{subfigure}
\hfill
\begin{subfigure}{0.48\textwidth}
\includegraphics[width=\textwidth]{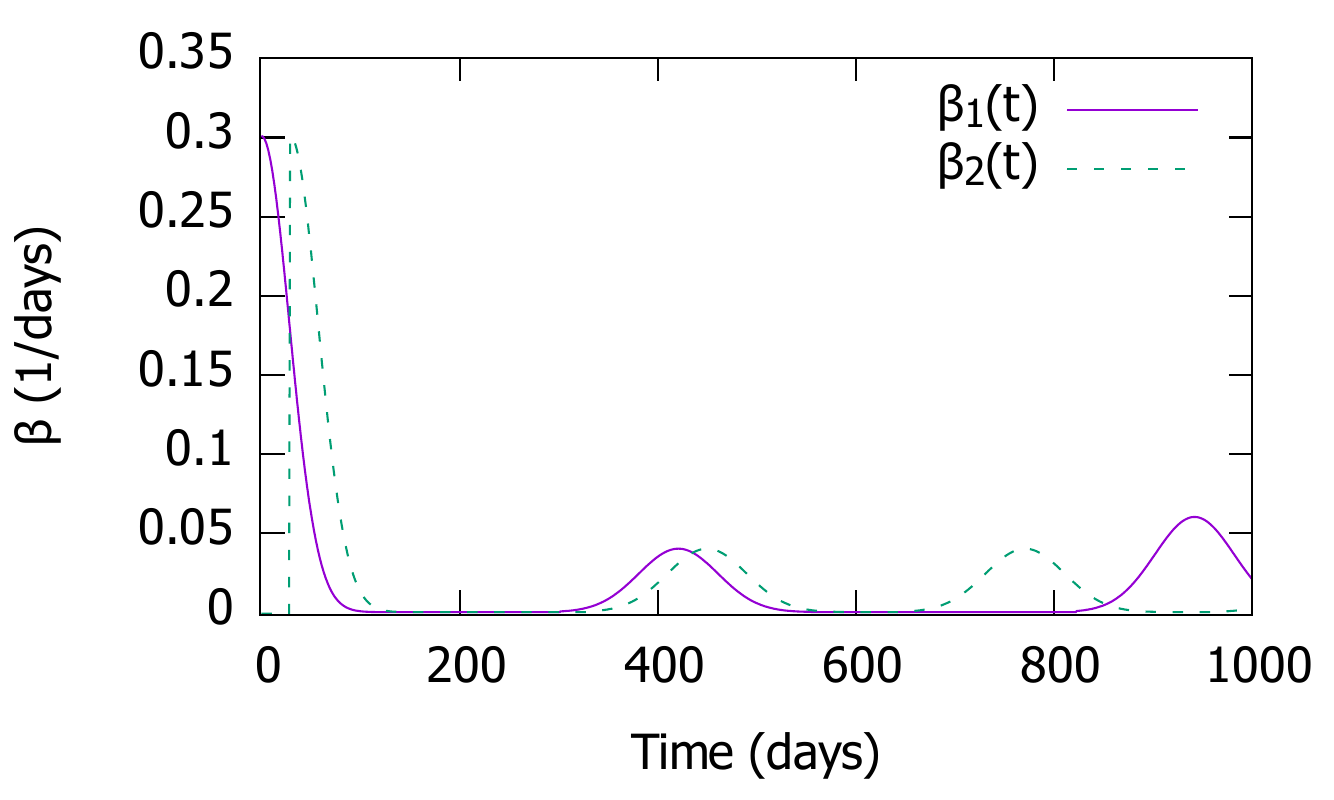}
\caption{The infection rate v.s. time with vaccine allocated on day 1000.}
\label{fig:beta(na0.04ex0.06c10alpha1000)}
\end{subfigure}
\hfill
\begin{subfigure}{0.48\textwidth}
\includegraphics[width=\textwidth]{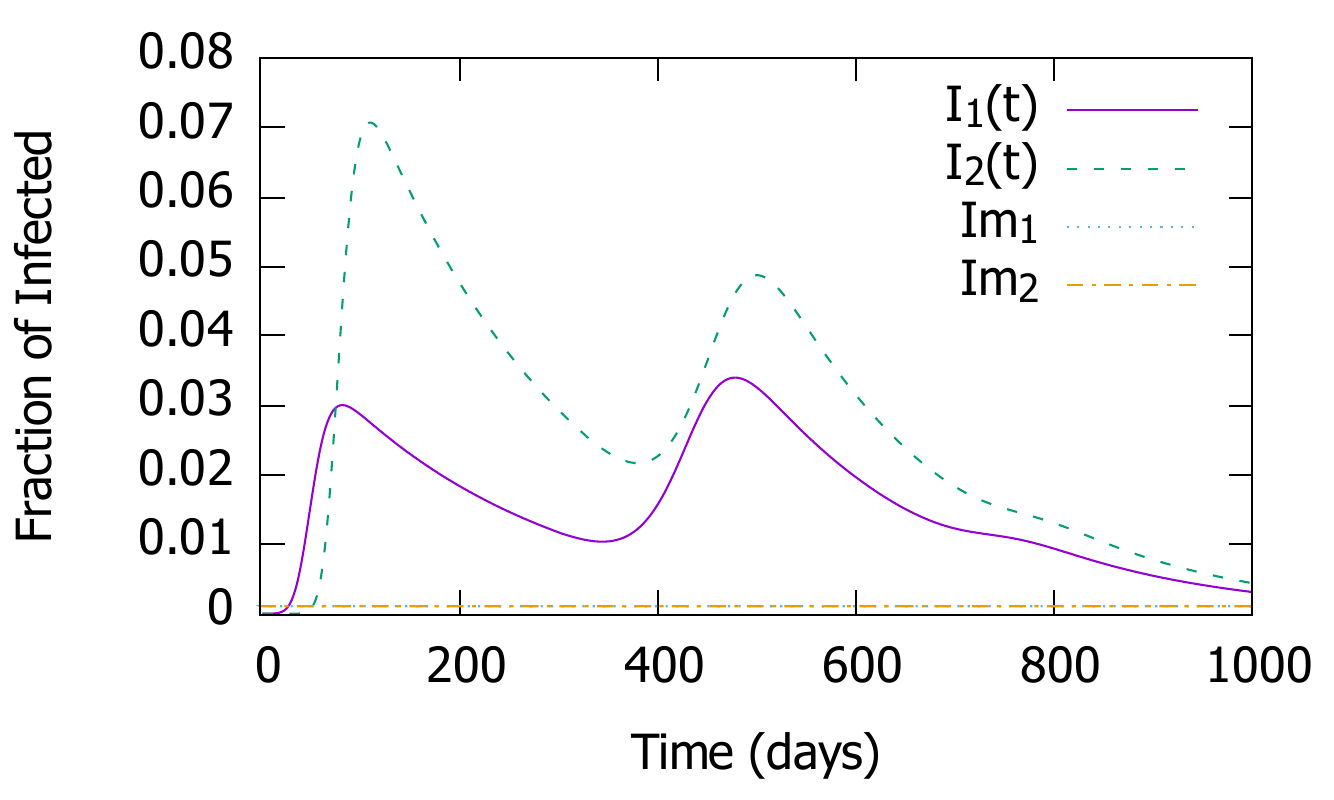}
\caption{The infected ratio v.s. time with vaccine allocated on day 300.}
\label{fig:i(na0.04ex0.06c10alpha300)}
\end{subfigure}
\hfill
\begin{subfigure}{0.48\textwidth}
\includegraphics[width=\textwidth]{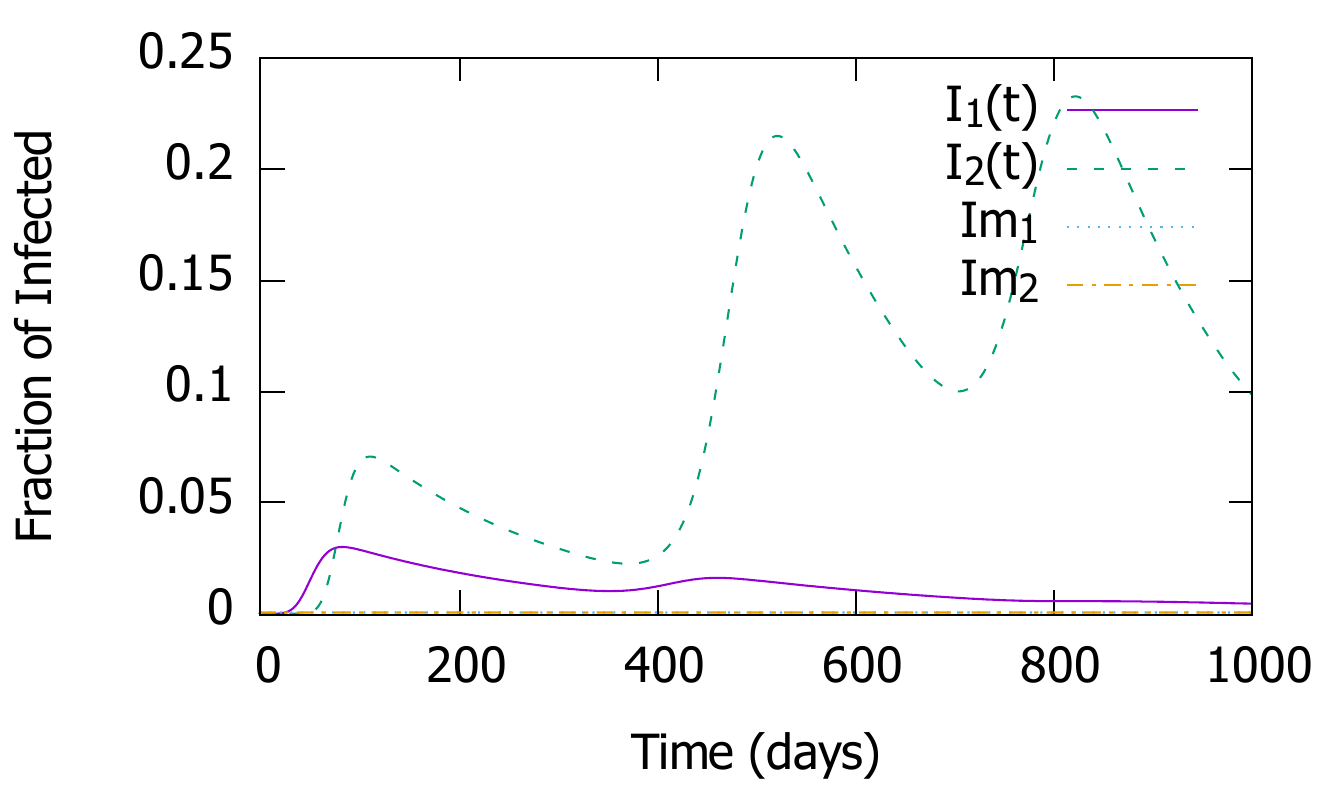}
\caption{The infected ratio v.s. time with vaccine allocated on day 1000.}
\label{fig:i(na0.04ex0.06c10alpha1000)}
\end{subfigure}
\caption{The infection rate and corresponding infected population ratio versus time for two different vaccine allocation time $\alpha=300$ and $\alpha=1000$. Other than the first peak in the infection rate, the infection rate due to virus variant from local sources is 0.04 and foreign ones is 0.06 with migration rate $c_1=c_2=1.42\times 10^{-4}$, and $c_3=7\times 10^{-6}$. $I_{m1}=I_{m2}=10^{-3}$ in both Fig.\ref{fig:i(na0.04ex0.06c10alpha300)} and Fig.\ref{fig:i(na0.04ex0.06c10alpha1000)}. $I_{m1}$, $I_{m2}$ are significantly smaller than both $I_1(t)$ and $I_2(t)$, and the line in Fig.\ref{fig:i(na0.04ex0.06c10alpha300)} or Fig.\ref{fig:i(na0.04ex0.06c10alpha1000)} is positioned very close to the horizontal axis.}
\label{fig:na0.04ex0.06c10}
\end{figure}

In this scenario, the infection rate $\beta_i(t)$ for country $i$ is modeled by a series of Gaussian functions centered at different times, along with a constant background value set at $0.001$. As depicted in Fig.\ref{fig:beta(na0.04ex0.06c10alpha300)} or Fig.\ref{fig:beta(na0.04ex0.06c10alpha1000)}, the onset of the pandemic for country 1 (at $t\simeq 0$) is represented by $\beta_1(t)$, the infection rate of country 1, which exhibits the shape of the right half of a Gaussian distribution. It has an amplitude of $0.3$ and a half-width of $30$ days around the peak, mirroring the initial peak seen in the Fig.\ref{fig:betausa1}. For country 2, the beginning of the pandemic is triggered by its population exchange with country 1. In other words, the first wave of the pandemic in country 2 is attributed to infected individuals originating from country 1. The epidemic in country 2 emerges when the infected population ratio surpasses $10^{-6}$, with the initial peak in the infection rate $\beta_2(t)$ exhibiting the same shape as that in $\beta_1(t)$. The migration rates depicted in Fig.\ref{fig:na0.04ex0.06c10} are denoted as $c_1=c_2=1.42\times 10^{-4}$ and $c_3=7\times 10^{-6}$. These values are ten times larger compared to the migration rates mentioned in subsection \ref{foursce}, illustrating a stronger connectivity between the two hypothetical countries in this context. Other parameters remain consistent with those cited in section \ref{modelp}, and $N_1=N_2=1$ in this subsection. For subsequent peaks of the infection rate $\beta_i(t)$, Gaussian functions with amplitudes of $0.04$ and $0.06$, respectively, and widths around 100 days are employed to describe the infection rates caused by domestic and imported virus variants.\\

Similar to subsection \ref{foursce}, country 1 initiates the vaccination of its own population starting on day 300. The decision on when to share half of the vaccination rate with country 2 is left to country 1. In an extremely benevolent scenario where country 1 opts to share from the outset (on day 300), the infection rate $\beta_i(t)$ versus time in Fig.\ref{fig:beta(na0.04ex0.06c10alpha300)} reveals two new native waves in both countries.\\

Conversely, in a more selfish scenario where country 1 delays vaccine distribution until day 1000, Fig.\ref{fig:beta(na0.04ex0.06c10alpha1000)} shows one native and one exotic wave in country 1, along with two native waves in country 2. The corresponding infected population ratios for both cases are plotted in Fig.\ref{fig:i(na0.04ex0.06c10alpha300)} and Fig.\ref{fig:i(na0.04ex0.06c10alpha1000)}. Prior to $t=300$, both figures exhibit identical curves. However, after $t=300$, the infected population ratio for country 2 in Fig.\ref{fig:i(na0.04ex0.06c10alpha1000)} is significantly larger than that in Fig.\ref{fig:i(na0.04ex0.06c10alpha300)} (approximately four times larger at the second peak and nearly ten times larger at the third peak). For country 1, Fig.\ref{fig:i(na0.04ex0.06c10alpha300)} only displays a slight decrease in its infected population ratio compared to Fig.\ref{fig:i(na0.04ex0.06c10alpha1000)}. The benefit of avoiding the third native wave is outweighed by the incoming third wave of exotic origin, as illustrated in Fig.\ref{fig:beta(na0.04ex0.06c10alpha1000)} and Fig.\ref{fig:i(na0.04ex0.06c10alpha1000)}. This once again highlights that while country 1 need not be excessively benevolent in sharing its vaccine resources from the very beginning, it should also avoid delaying vaccine sharing too late for its own benefit.\\ 

\begin{figure}[h]
\centering
\begin{subfigure}{0.48\textwidth}
\includegraphics[width=\textwidth]{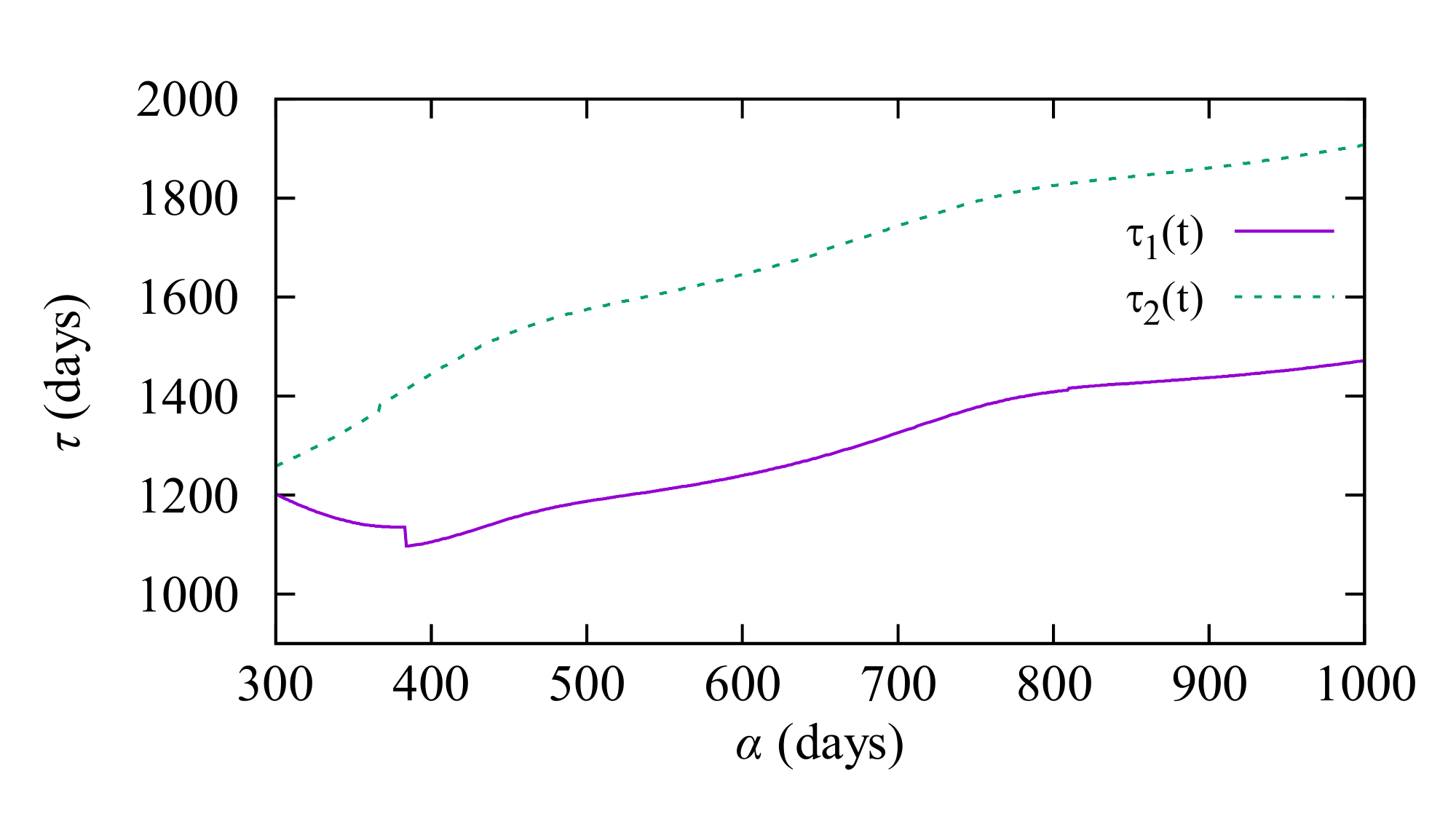}
\caption{The infection rate due to virus variant from local sources is 0.04 and foreign ones is 0.06 with migration rate$c_1=c_2=1.42\times 10^{-4}$ and $c_3=7\times 10^{-6}$.}
\label{fig:tau(na0.04ex0.06c10)}
\end{subfigure}
\hfill
\begin{subfigure}{0.48\textwidth}
\includegraphics[width=\textwidth]{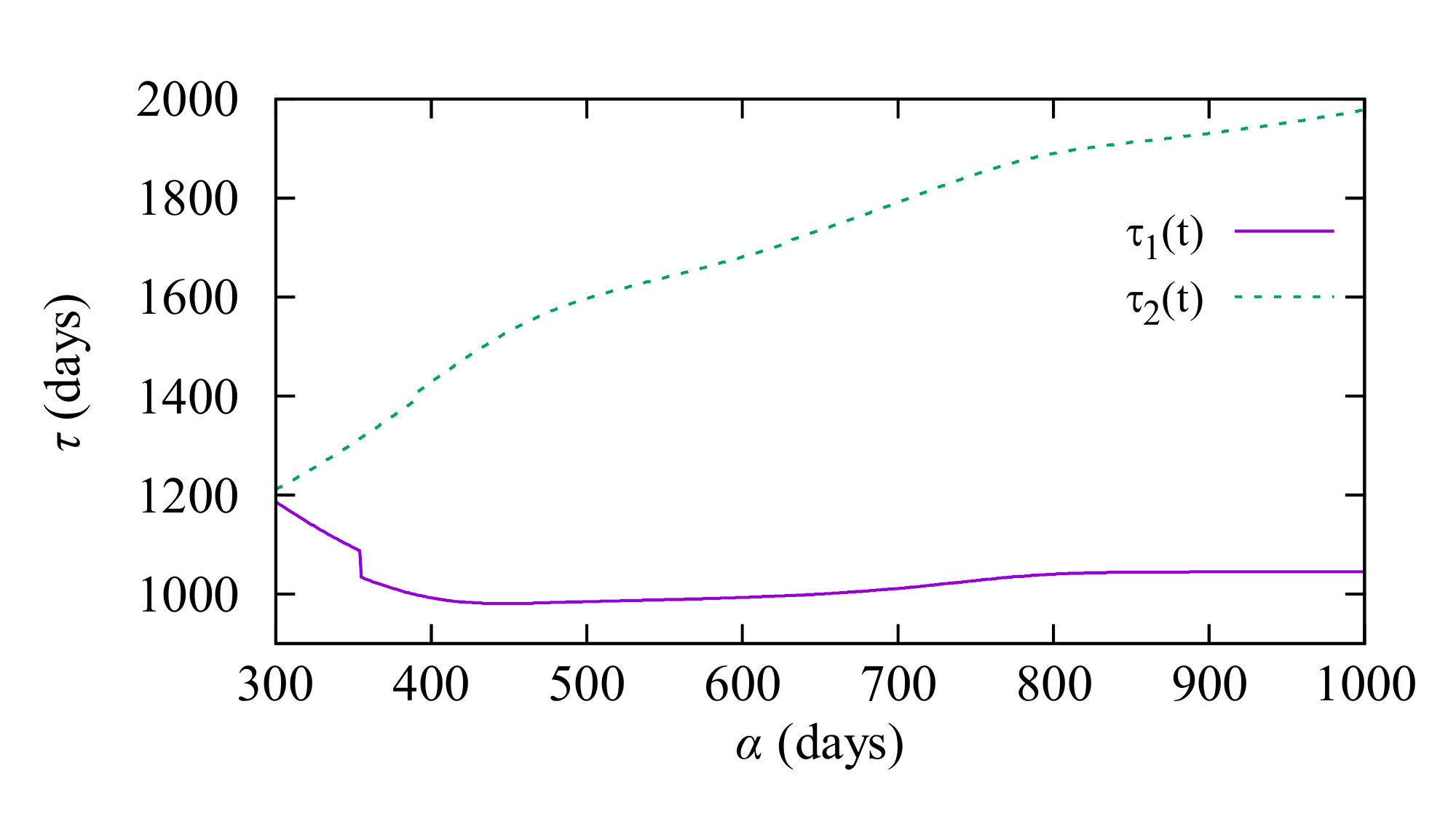}
\caption{The infection rate due to virus variant from local sources is 0.04 and foreign ones is 0.06 with migration rate$c_1=c_2=1.42\times 10^{-5}$ and $c_3=7\times 10^{-7}$.}
\label{fig:tau(na0.04ex0.06c1)}
\end{subfigure}
\hfill
\begin{subfigure}{0.48\textwidth}
\includegraphics[width=\textwidth]{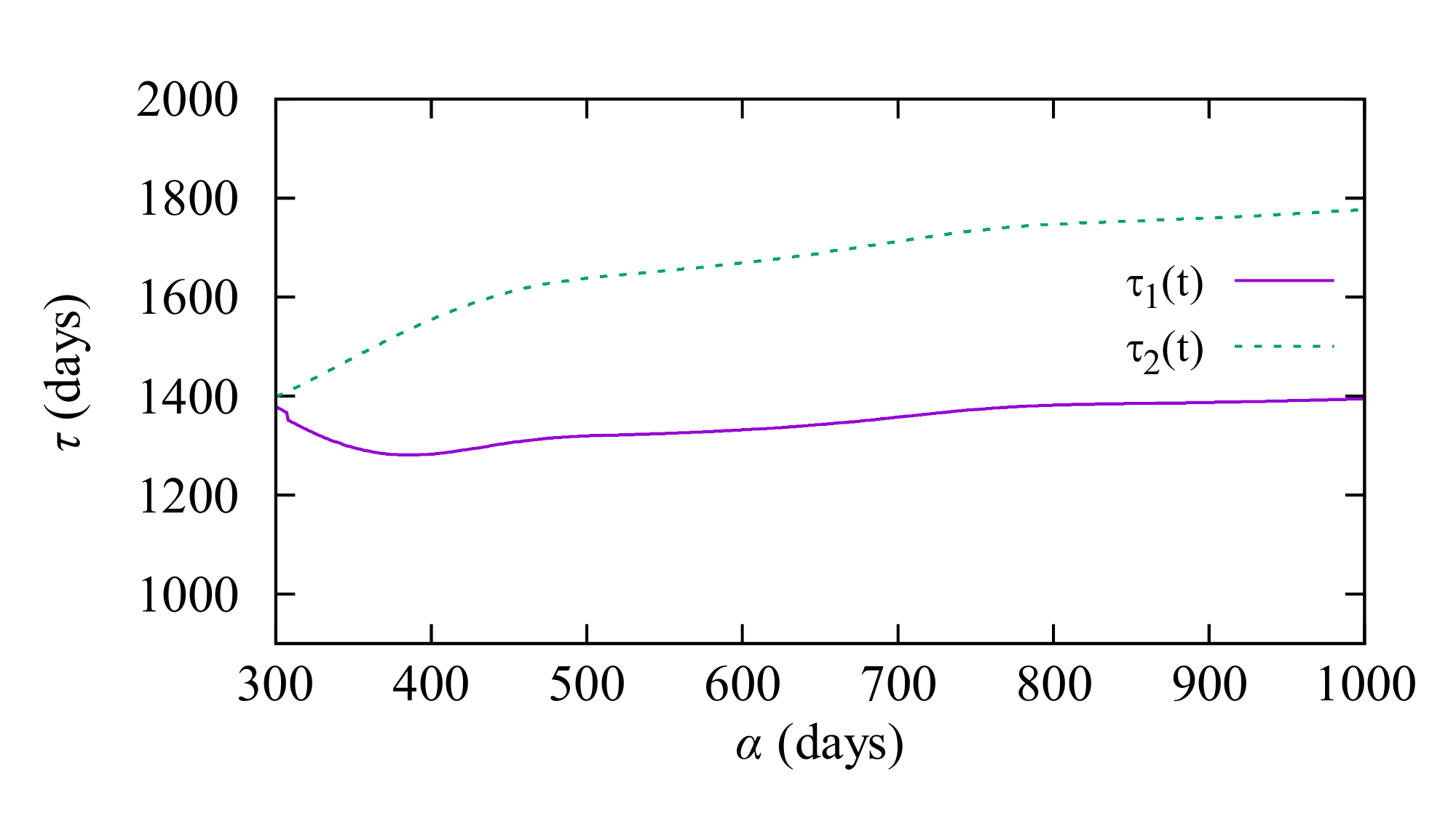}
\caption{The infection rate due to virus variant from local sources is 0.06 and foreign ones is 0.04 with migration rate$c_1=c_2=1.42\times 10^{-4}$ and $c_3=7\times 10^{-6}$.}
\label{fig:tau(na0.06ex0.04c10)}
\end{subfigure}
\hfill
\begin{subfigure}{0.48\textwidth}
\includegraphics[width=\textwidth]{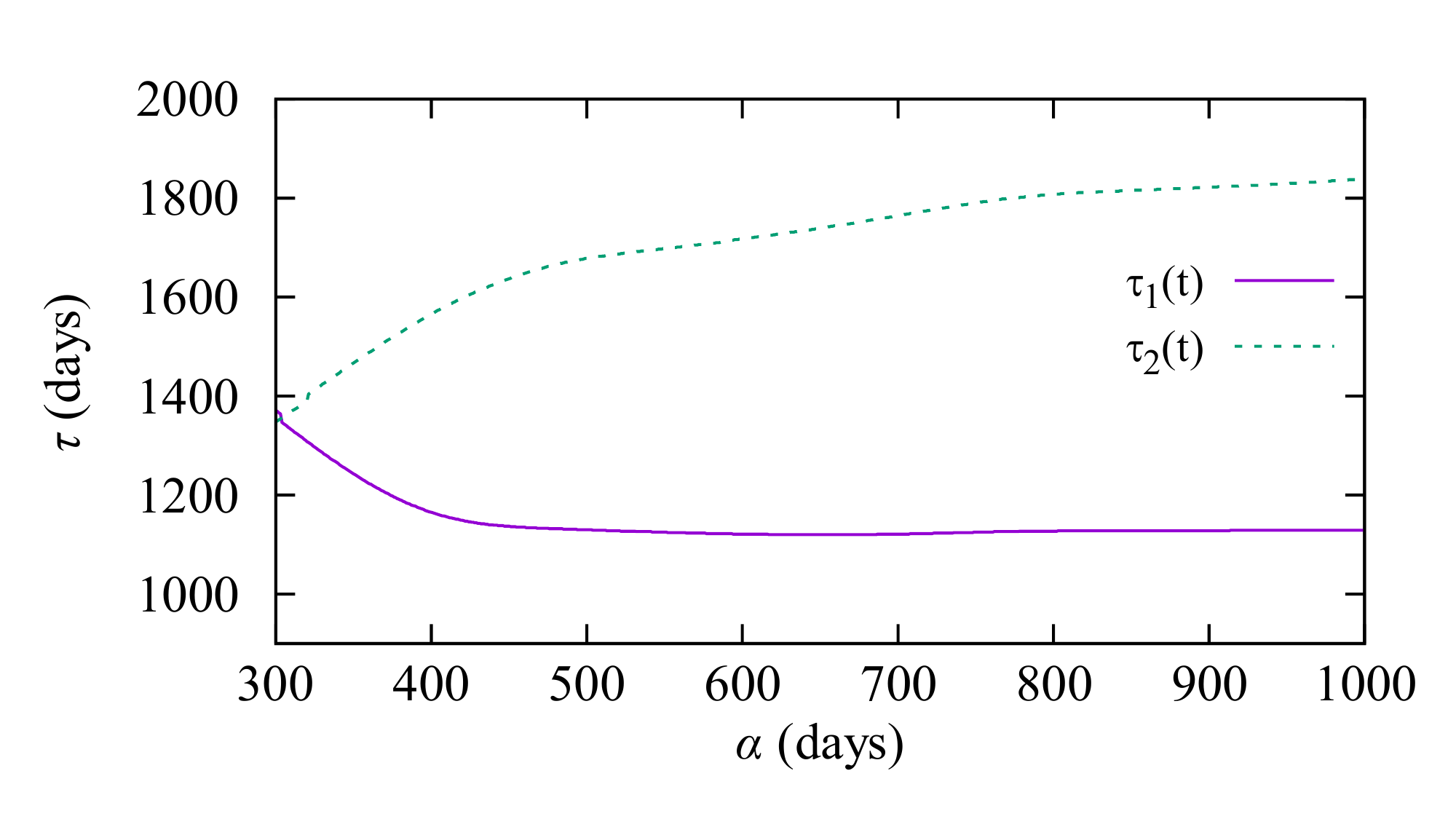}
\caption{The infection rate due to virus variant from local sources is 0.06 and foreign ones is 0.04 with migration rate$c_1=c_2=1.42\times 10^{-5}$ and $c_3=7\times 10^{-7}$.}
\label{fig:tau(na0.06ex0.04c1)}
\end{subfigure}
\caption{The epidemic control time $\tau$ versus vaccine allocation time $\alpha$ for different infection rate due to virus variants and migration rate.}
\label{fig:taunaexc}
\end{figure}

To determine the optimal timing for distributing vaccines for country 1, we plot the epidemic control time $\tau_i$ as a function of vaccine distribution time $\alpha$ for both countries in Fig.\ref{fig:tau(na0.04ex0.06c10)}. The overall trajectories for these two curves resemble those shown in Fig.\ref{fig:tau1tau24tua14tau2}. However, a notable difference from Fig.\ref{fig:tau1tau24tua14tau2} is the sharp dip at $\alpha=382$ for country 1, with almost identical positive slopes for both countries after this dip. The dip observed for country 1 in Fig.\ref{fig:tau(na0.04ex0.06c10)} arises from the presence of two native variants if vaccines are distributed before day 382 (as depicted in the case for $\alpha=300$ in Fig.\ref{fig:beta(na0.04ex0.06c10alpha300)}), whereas there is only one native variant if vaccines are allocated after day 382 (as shown in the case for $\alpha=1000$ in Fig.\ref{fig:beta(na0.04ex0.06c10alpha1000)}, where the last peak is from an exotic virus variant).In this scenario, the optimal timing for country 1 to share vaccines with country 2 is around day 383, approximately two to three months after the extensive vaccination campaign in country 1.\\

The almost identical slope observed after $\alpha=382$ is attributed to the interconnectivity of the two countries, which is controlled by the migration terms $c_i$ in our model. If we adjust the migration terms ci to the same magnitude as those in subsection \ref{foursce}, the disparity in epidemic control time between the two countries increases with the allocation time , as illustrated in Fig.\ref{fig:tau(na0.04ex0.06c1)}. In Fig.\ref{fig:tau(na0.04ex0.06c1)}, the positive slope of country 1 after the minimum becomes smaller, and the dip shifts to smaller values. This shift indicates that country 1 does not need to withhold its vaccine resources for as long to avoid the native variant. Both the slope and the change in the dip location suggest that more frequent contacts result in a stronger correlation in the evolution of the pandemic in both countries.\\

From the data\cite{91COVID}, it's not evident that the infection rate of native virus variants would necessarily be smaller than that of exotic virus variants. Thus, we explore another possibility here by swapping the amplitudes of the Gaussian functions in the infection rate $\beta_i(t)$ for native and exotic virus variants. Specifically, we set the peak value of the Gaussian function to be $0.06$ for native virus variants and $0.04$ for exotic virus variants. This setup results in the infected ratio caused by native virus variants being larger in both countries, making it more challenging to meet the conditions for infection caused by exotic virus variants.\\ 

Remarkably, we find that there are no exotic virus variants in this scenario. The epidemic control time varies more smoothly with the vaccine allocation time, and the slopes of the curves for both countries are smaller compared to previous cases (see Fig.\ref{fig:taunaexc}). The smaller slope indicates a more severe epidemic situation in both countries. The small dip located nearby $\alpha\simeq 300$ in Fig.\ref{fig:tau(na0.06ex0.04c10)} is related to the third wave of native virus variants, which would vanish if $\alpha$ were greater than around 310. Once again, from Fig.\ref{fig:tau(na0.06ex0.04c10)} and Fig.\ref{fig:tau(na0.06ex0.04c1)}, we observe that country 1 should distribute its vaccines no later than day 380 to 400 to minimize its own epidemic control time $\tau_1$, although the benefit for country 1 may not be as evident with smaller connectivity with country 2.\\

\section{Conclusion}\label{conclusion}
One significant obstacle to achieving equitable access to COVID-19 vaccines is the race among high-income countries to secure vaccines early in the mass vaccination stage\cite{Khairi}. This pursuit of self-interest by individual countries is perhaps unavoidable and could impede international efforts toward equitable access to medical resources for future epidemics. However, we aim to demonstrate that sharing vaccines and pursuing self-interest are not necessarily mutually exclusive.\\ 

To illustrate this point, we employ two coupled SIRD models, with parameters fitted using real data from the USA, to examine the relationship between the epidemic control time $\tau_i$ of two hypothetical countries and the timing of vaccine distribution $\alpha$, where the vaccines are owned by one of the countries. We also consider the possibility of emergent virus variants as the epidemic progresses, incorporating a series of Gaussian functions into the infection rate $\beta_i(t)$.\\

In all cases, our findings suggest that the country possessing the vaccines (representing high-income countries) should distribute them no later than a certain point in time. The optimal timing for distributing vaccines for this country is determined by the interconnectivity between the two countries and its own pandemic situation. Interestingly, we observe that the effect of providing more vaccines at a later time is roughly equivalent to providing fewer vaccines at an earlier time. Conversely, for the country receiving the vaccines (representing low to middle-income countries), the best timing for vaccine distribution is always on the first available day ($\alpha=300$ in previous discussions), which may not align with the best timing for the country distributing the vaccines.\\

We have also set other parameters such as recovery rate and death rate to be identical for both countries to facilitate a controlled comparison. However, in reality, these parameters would vary among countries with different economic capabilities, potentially resulting in higher human costs for low to middle-income countries.\\

Our main findings align with those from more complex analyses\cite{Ye,Moore,Gozzi,Li,Ning,Bayati}, which generally require a larger amount of data and input parameters. Specifically, we have explicitly evaluated the epidemic control time $\tau_i$ as a function of vaccine distribution time $\alpha$ using a model with simpler mathematical formulas and parameters more directly linked to available data. For instance, most compartmental models utilized by the studies cited in this paper include the exposed population (referred to as the SEIRD model), which often necessitates estimation based on an average incubation period, a parameter that may lack robust data.\\

Another advantage of this simplified model used for the COVID-19 pandemic here is its adaptability to describe systems beyond epidemics. For instance, the strategy of controlling desert locusts by deploying biopesticides in various countries or regions shares a conceptual similarity with using vaccines to halt the spread of viruses. Furthermore, expanding the model to include additional countries or regions can be done with ease, rendering it highly versatile for addressing a diverse array of applications.\\

 In contrast, the model employed for the COVID-19 pandemic in this paper is easily modifiable to describe other systems beyond epidemics. For example, the control of desert locusts by applying biopesticides in different countries or regions bears a strong resemblance to the concept of using vaccines to prevent the spread of viruses. Additionally, incorporating more countries or regions into the model can be accomplished straightforwardly, making it versatile for a wide range of applications.\\

Certainly, reminding high-income countries about the threat of new virus strains\cite{Ye} and emphasizing the importance of promptly sharing vaccines do not entirely resolve the dire pandemic situations faced by many low to middle-income countries. Certain Non-Pharmaceutical Interventions (NPIs) tailored for each individual country are still necessary\cite{Mukaigawara}, even with the assistance of vaccination, to mitigate the pandemic. Additionally, it is crucial to carefully select the vaccination program to maximize the chances of halting the virus outbreak in different countries\cite{Toledano}. Addressing intellectual property issues\cite{Pilkington} and allowing pharmaceutical companies to profitably produce and sell vaccines in low to middle-income countries\cite{Du} may also help alleviate the problems of inequitable access to vaccines. 
   
\section*{Acknowledgment}
S. P. Chao acknowledges the financial support from MOST in Taiwan (Grant No.110-2112-M-017-001) during the early stage of this work. 


\begin{thebibliography}{}
\bibitem{ft1}The normalized inbound arrival is about $1.58\times 10^{-4}$. The $c_1$ value cited is further divided by $11$ to account for different countries (as the top two countries, Canada and Mexico taking much more quantities in the top ten countries). The $c_3$ is related to the infected from other country making contact with its own population and we set $c_3=0.05c_1$.
\bibitem{footnote1} The vaccination rate $\nu(t)=1.5\times 10^{-3}\Theta(t-300)$ is estimated by $0.5$ million vaccine doses administered per day in average\cite{91COVID} from $t=300$ to $t=700$.
\end{thebibliography}
\end{document}